\documentclass[a4paper,12pt,final]{article}
\def\figformat{png}
\pdfoutput=1
\usepackage{geometry}
\usepackage{calc}
\usepackage[T1]{fontenc}
\usepackage{ifthen}
\usepackage{amsmath}
\usepackage{amsfonts}
\usepackage{amsthm}
\usepackage{amssymb}
\usepackage{multirow}
\usepackage{enumerate}
\usepackage{dsfont}
\usepackage[all]{xy}
\usepackage{epsfig}
\usepackage{color}
\usepackage{hyperref}
\definecolor{gray}{gray}{.5}
\newtheorem{Lem}{Lemma}[section]
\newtheorem{Pro}[Lem]{Proposition}
\newtheorem{Thm}[Lem]{Theorem}
\newtheorem{Cor}[Lem]{Corollary}
\theoremstyle{definition}
\newtheorem{Def}[Lem]{Definition}
\newtheorem{Exa}[Lem]{Example}
\newtheorem{Rem}[Lem]{Remark}
\newtheorem*{Ack}{Acknowledgements}
\newcommand{\bA}{\mathds{A}}
\newcommand{\bC}{\mathds{C}}
\newcommand{\bE}{\mathds{E}}

\newcommand{\bN}{\mathds{N}}
\newcommand{\bQ}{\mathds{Q}}
\newcommand{\bR}{\mathds{R}}
\newcommand{\cA}{\mathcal A}
\newcommand{\cB}{\mathcal{B}}
\newcommand{\cC}{\mathcal C}
\newcommand{\cE}{\mathcal E}
\newcommand{\cL}{\mathcal L}
\newcommand{\cM}{\mathcal M}

\newcommand{\cP}{\mathcal P}
\newcommand{\id}{\mathds{1}}
\newcommand{\tr}{{\rm tr}}
\newcommand{\sa}{{\cA_{\rm sa}}}
\newcommand{\aff}{{\rm aff}}
\newcommand{\ri}{{\rm ri}}
\newcommand{\rb}{{\rm rb}}
\newcommand{\I}{{\it I}}
\newcommand{\rI}{{\it rI}}
\newcommand{\cl}{{\rm cl}_\rI}
\newcommand{\de}{{\rm d}}
\def\conv{\mathop{\text{conv}}}
\def\argmax{\mathop{\text{argmax}}}

\def\ext{\mathop{\text{ext}(\cE)}}
\begin{document}
\thispagestyle{empty}
%
%
%
%
\begin{center} 
\textbf{\Large{Continuity of the Maximum-Entropy Inference}}\\
\vspace{.3cm}
Stephan Weis\\
Max-Planck-Institute for Mathematics in The Sciences\\
Leipzig, Germany\\
\texttt{maths@weis-stephan.de}\\[1mm]
April 21, 2014
\end{center}
%
%
%
%
\begin{abstract}
We study the inverse problem of inferring the state of a finite-level 
quantum system from expected values of a fixed set of observables, by 
maximizing a continuous ranking function. We have proved earlier that
the maximum-entropy inference can be a discontinuous map from the 
convex set of expected values to the convex set of states because the
image contains states of reduced support, while this map restricts to a 
smooth parametrization of a Gibbsian family of fully supported states. 
Here we prove for arbitrary ranking functions that the inference is 
continuous up to boundary points. This follows from a continuity 
condition in terms of the openness of the restricted linear map from 
states to their expected values. The openness condition shows also that 
ranking functions with a discontinuous inference are typical. Moreover 
it shows that the inference is continuous in the restriction to any 
polytope which implies that a discontinuity belongs to the quantum 
domain of non-commutative observables and that a geodesic closure of a 
Gibbsian family equals the set of maximum-entropy states. We discuss 
eight descriptions of the set of maximum-entropy states with proofs of 
accuracy and an analysis of deviations.
\end{abstract}
\section{Introduction}
\label{sec:intro}
\par
Boltzmann's pioneering derivation of the maximum-entropy principle 
\cite{Boltzmann77} is, isolated from its origins in statistical physics, 
a counting problem and can be solved asymptotically using Stirling's 
approximation of factorials. This derivation is a welcome topic
in the literature of the history of science, see for example Uffink 
\cite{Uffink06} Section~4.4, and in textbooks of physics, see for 
example Caticha \cite{Caticha12} Section~3.6. In 1957 Jaynes 
\cite{Jaynes} has highlighted the independence of the maximum-entropy 
principle from the physical context and has seen it as a universal tool 
for inference, while supporting his claim with Shannon's axioms of 
uncertainty \cite{Shannon48}. Since then the maximum-entropy principle 
has become an ubiqui\-tous method in science as for example 
the {\it International Workshop on Bayesian Inference and Maximum Entropy 
Methods in Science and Engineering} continues to demonstrate. 
\par
A better argument supporting the maximum-entropy principle as a tool of 
inference was found in 1980 by Shore and Johnson \cite{Shore-Johnson80}. They 
have derived the {\it minimum discrimination information principle} 
\cite{Csiszar75,Topsoe} or {\it ME method} \cite{Caticha-Giffin06}, 
formulated earlier by Kullback, from axioms of inference rather than axioms 
of uncertainty. For simplicity we consider a finite state space $X$. The ME 
method updates a prior probability measure $Q$ on $X$, when new information 
becomes available. New information is assumed in the
form of a constraint, that is a subset of probability measures on $X$, and 
the prior $Q$ is updated to the probability measure $P$ in the constraint 
set which minimizes the {\it relative entropy}
\[
D(P\|Q):=\int_X\log(\frac{{\rm d}P}{{\rm d}Q}){\rm d}P
\]
from $Q$. Thereby $\frac{{\rm d}P}{{\rm d}Q}$ is the Radon-Nikod\'ym 
derivative of $P$ with respect to $Q$ if $P$ is absolutely continuous with 
respect to $Q$. Otherwise $D(P\|Q)=\infty$. If the prior $Q$ is uniform 
then the ME method reduces to the maximum-entropy principle. The axioms of 
the ME method are still being discussed. Literature supporting the ME method 
includes Skilling, Csisz\'ar or Caticha and Giffin 
\cite{Skilling88,Csiszar91,Caticha-Giffin06} while critics such as Karbelkar 
or Uffink \cite{Karbelkar86,Uffink95} accept the ME method for a composite 
system only if the subsystems are uncorrelated.
\par
In quantum mechanics, a joint distribution and other notions of probability
theory are problematic, as for example Davies and Lewis \cite{Davies-Lewis70} 
have pointed out. This may be one of the reasons why an axiomatic approach
to a quantum analogue of the ME method is still missing, although the problem 
is of interest, as Ali et al.\ \cite{Ali12} have shown. Nevertheless, von 
Neumann \cite{vonNeumann27} has formally generalized the maximum-entropy 
principle already in 1927 to quantum states. The canonical states which arise
from linear constraints are central in quantum statistical mechanics, see 
Bratteli and Robinson \cite{Bratteli}. Von Neumann's maximum-entropy principle 
has also been considered in quantum estimation, where Bu\v zek et al.\ 
\cite{Buzek99} have compared it to other estimation methods. Ruskai 
\cite{Ruskai88} has analyzed the quantum analogue of the ME method for linear
constraints in an infinite-dimensional setting.
\par
We have discovered a topological problem of quantum inference under linear 
constraints already in finite dimensions, namely its discontinuity 
\cite{Weis-Diss,Weis-Knauf}. We will show that the discontinuity follows from
the convex geometry of the space of quantum states. As we will see, almost all 
methods of quantum inference from expected values have discontinuities but 
this happens only if the observables do not commute. So the discontinuity is a 
pure quantum effect, like for example entanglement described by Schr\"odinger
\cite{Schroedinger35,Nielsen-Chuang,Greenberger09}. In contrast to entanglement
we think a discontinuous inference is not experimentally measurable and 
therefore has no consequences in physics, see the end of the section. The aim 
of the present article is to continue our work \cite{Weis-Knauf,Weis-topology} 
and to integrate it into a larger mathematical context.
\par
We sketch the discontinuity problem starting from the continuous setting 
of probability distributions. The constraints are defined by intersecting the 
probability simplex $\cP$ of probability measures 
on $X$ with the fibers of expected values of some measurable functions 
$f_i:X\to\bR$, $i=1,\ldots,k$. If these functions ${\bf f}:=(f_1,\ldots,f_k)$ 
and a prior $Q$ are fixed, then the ME method determines a mapping from $k$-tuples 
of expected values to probability distributions,
\addtocounter{equation}{-1}
\begin{equation}\label{eq:continuous-map}
\{\int_X{\bf f}{\rm d}P \in \bR^k \mid P\in\cP\}
\quad\to\quad
\cP.
\tag{*}
\end{equation}
\addtocounter{equation}{+1}%
It is well-known that (\ref{eq:continuous-map}) restricts to the smooth map
\addtocounter{equation}{-1}
\begin{equation}\label{eq:real-analytic}
\{\int_X{\bf f}{\rm d}P \in \bR^k
\mid P\in\cP, \text{$P$ has full support $X$}\}
\quad\to\quad
\cP,
\tag{**}
\end{equation}
\addtocounter{equation}{+1}%
whose image is the {\it exponential family} $\{P(\gamma)\mid\gamma\in\bR^k\}$
with densities $\tfrac{{\rm d}P(\gamma)}{{\rm d}Q}
=e^{\gamma_1f_1+\cdots+\gamma_kf_k-\Lambda(\gamma)}$,
$\gamma=(\gamma_1,\ldots,\gamma_k)\in\bR^k$ and $\Lambda:\bR^k\to\bR$ is for 
normalization. Barndorff-Nielsen has shown in \cite{Barndorff}, p.~154, that 
(\ref{eq:real-analytic}) always extends to the continuous map 
(\ref{eq:continuous-map}). Csisz\'ar \cite{Csiszar91} anticipates in his axioms 
of inference the continuity of (\ref{eq:real-analytic}), but not the continuity 
of (\ref{eq:continuous-map}). The analogue of (\ref{eq:real-analytic}) for 
finite-level quantum systems is real analytic but a continuous extension can be 
missing.
\par
Mathematically, the state of a finite-level quantum system is represented by a 
density matrix in a (complex) C*-algebra of complex square matrices 
${\rm Mat}(n,\bC)$, of size $n\in\bN$. To have a three-dimensional example of a 
discontinuity we generalize to a {\it real} 
C*-subalgebra $\cA\subset{\rm Mat}(n,\bC)$. The simplest example in a complex 
algebra has four dimensions. See Section~\ref{sec:real-algebras} for references
and examples.
\begin{Def}[Inference]\label{def:entropic-inference}
The real vector space $\sa$ of self-adjoint matrices in $\cA$ is a Euclidean 
space with the restricted Hilbert-Schmidt inner product 
$\langle a,b\rangle:=\tr(ab^*)$, $a,b\in\cA$. The {\it state space} of $\cA$ 
is defined by
\[
\cM=\cM_\cA:=\{\rho\in\cA\mid\rho\succeq 0,\tr(\rho)=1\},
\]
where $\rho\succeq 0$ means $\rho$ is positive semi-definite and $\tr(\rho)$
is the sum of diagonal elements of $\rho$. We use the terms of 
{\it density matrix} and {\it state} synonymously for a matrix in $\cM$. 
We usually keep a sequence $u_1,\ldots,u_k\in\sa$, $k\in\bN$, of self-adjoint 
matrices, called {\it observables}, fixed. We put ${\bf u}:=(u_1,\ldots,u_k)$ 
and call the linear map 
\[
\bE=\bE_{\bf u}:\sa\to\bR^k,
\qquad
a\mapsto\langle a,u_i\rangle_{i=1}^k
\]
{\it expected value functional}. We call $\cC=\cC_{\bf u}:=\bE_{\bf u}(\cM)$ 
{\it convex support} in analogy to the name by which Barndorff-Nielsen,
Csisz\'ar and Mat\'u\v s \cite{Barndorff,CM03} call $\cC$ for probability 
distributions which, in the form of diagonal matrices, belong to our setting.  
\par
Let $\phi:\cM\to\bR$, the {\it ranking function}, be a continuous real valued 
function with a unique maximizer on each fiber $\{\rho\in\cM\mid\bE(\rho)=x\}$, 
$x\in\cC$. We define
\[\begin{array}{lrll}
\textit{maximum} 
& \psi:
& \cC\to\bR,
& x\mapsto\max\{\phi(\rho)\mid\bE(\rho)=x, \rho\in\cM\},\\
\textit{inference}
& \Psi:
& \cC\to\cM,
& x\mapsto\argmax\{\phi(\rho)\mid\bE(\rho)=x, \rho\in\cM\}.
\end{array}\]
Umegaki \cite{Umegaki} has defined the {\it relative entropy} $S(\rho,\sigma)$ 
of two states $\rho,\sigma\in\cM$ by
$S(\rho,\sigma):=\tr\,\rho(\log(\rho)-\log(\sigma))$ if the image of $\rho$ is 
included in the image of $\sigma$ if $\rho$ and $\sigma$ are considered linear 
maps in the same basis. Otherwise we set $S(\rho,\sigma):=+\infty$. Depending 
on a self-adjoint matrix $\theta\in\sa$ an invertible {\it prior state} is 
defined by $\sigma_\theta:=e^\theta/\tr(e^\theta)$, and we write 
$\phi_\theta(\rho):=-S(\rho,\sigma_\theta)$. The maximum of the ranking function
$\phi_\theta$ is denoted by $\psi_\theta$, the corresponding inference 
$\Psi_\theta$ is called {\it ME-inference}.
\end{Def}
\par
For completeness, we recall $S(\rho,\sigma)\geq 0$ for all $\rho,\sigma\in\cM$ 
and that $S(\rho,\sigma)=0$ implies $\rho=\sigma$, but $S$ is not a metric. The 
ME-inference $\Psi_\theta$ is well-defined because $\rho\mapsto-S(\rho,\sigma)$ 
is continuous and strictly concave when $\sigma$ is invertible. See for example 
Nielsen and Chuang or Wehrl \cite{Nielsen-Chuang,Wehrl} about these statements. 
The {\it von Neumann entropy} of $\rho\in\cM$ is $H(\rho):=-\tr\,\rho\log(\rho)$. 
We call $\Psi_0$ {\it maximum-entropy inference} because 
$H(\rho)=\log\tr(\id)-S(\rho,\id/\tr(\id))$. 
\begin{figure}
\begin{center}
\begin{picture}(13,5)
\ifthenelse{\equal{\figformat}{eps}}{%
\put(0,0){\includegraphics[height=5cm, bb=70 50 420 380, clip=]%
{kegel.eps}}
\put(8,0){\includegraphics[height=5cm, bb=0 0 500 519, clip=]%
{kegelbild.eps}}}{%
\put(0,0){\includegraphics[height=5cm, bb=294 210 1764 1596, clip=]%
{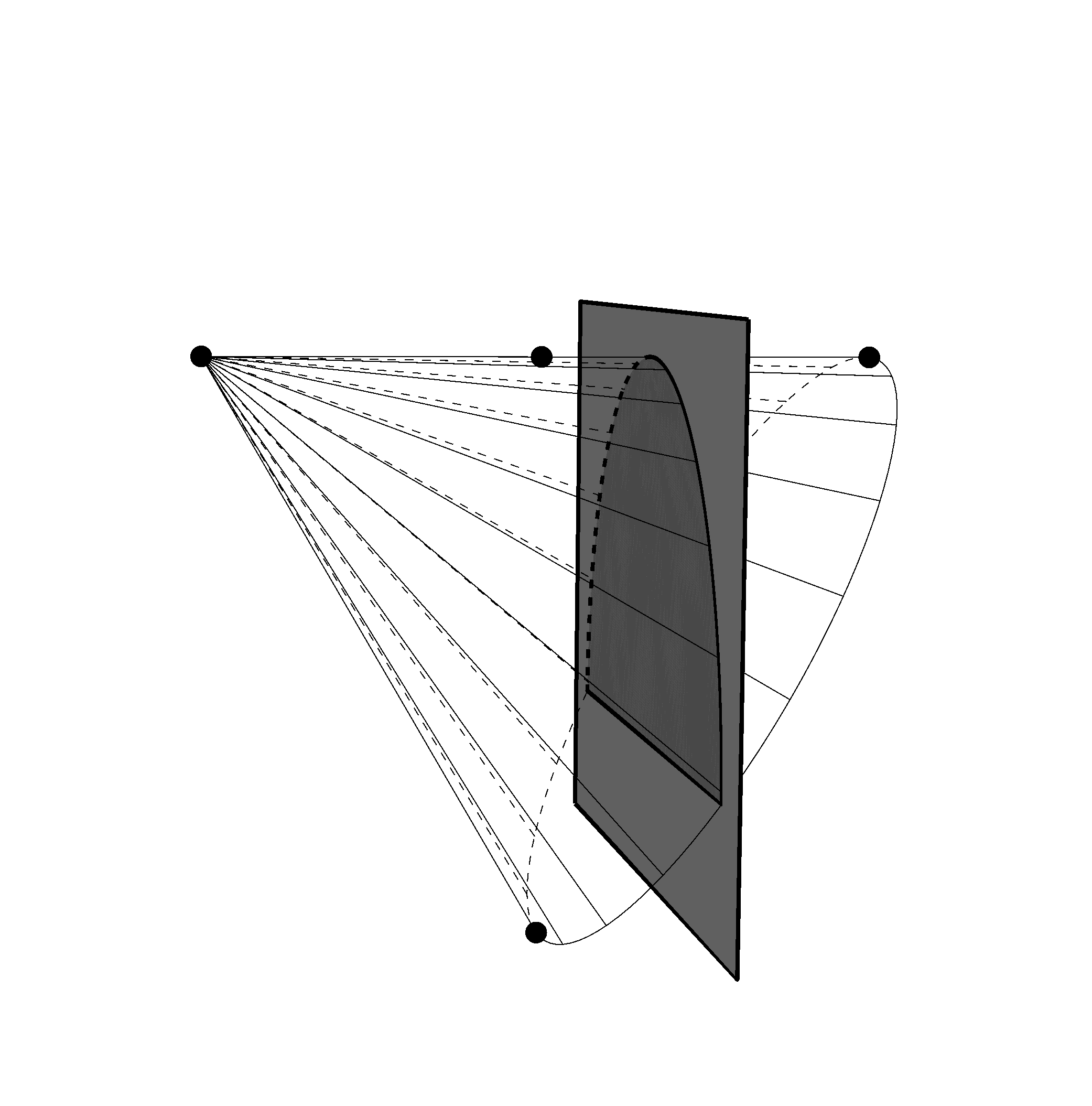}}
\put(8,0){\includegraphics[height=5cm, bb=0 0 500 517, clip=]%
{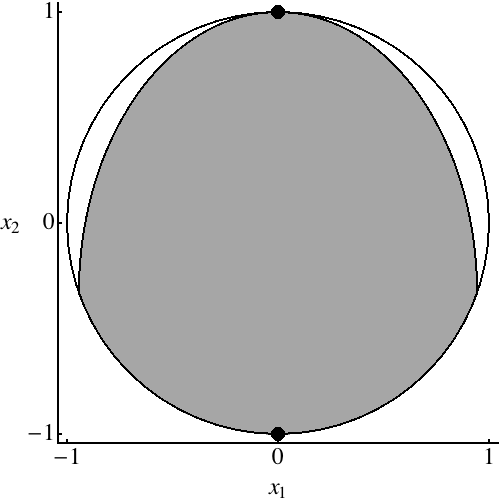}}}
\put(0,0){a)}
\put(0,4.6){$0_2\oplus 1$}
\put(2.5,4.6){$c$}
\put(4.7,4.6){$\rho(0)$}
\put(1.8,0.2){$\rho(\pi)$}
\put(8,0){b)}
\put(10.4,4.3){$\bE(c)$}
\put(11.3,2.2){$\bE(N)$}
\put(10.0,0.9){$\bE(\rho(\pi))$}
%
%
%
\end{picture}
\caption{\label{fig1}a) The cone state space $\cM^{\rm Cone}$. The depicted 
plane bounds a neighborhood $N$ of $c$. b) The unit disc 
$\cC=\bE(\cM^{\rm Cone})$. The image $\bE(N)$, in gray, is bounded by an 
ellipse of curvature $>1$ at $\bE(0_2\oplus 1)=\bE(c)=\bE(\rho(0))=(0,1)$. 
This shows that $\bE(N)$ is not a neighborhood of $\bE(c)$ in $\cC$, so 
$\bE|_{\cM^{\rm Cone}}$ is not open at $c$.}
\end{center}
\end{figure}
\par
To see why the inference $\Psi$ is not always continuous we recall that the
continuity of $\Psi$ at $x\in\cC$ means that any small perturbation of $x$ 
can be realized by an appropriate small perturbation of $\Psi(x)$ inside the 
set of inference states $\Psi(\cC)$. This implies that any small perturbation 
of $x$ can be realized by an appropriate small perturbation of $\Psi(x)$ 
inside the state space $\cM$. The latter condition is equivalent to the 
openness of the expected value functional $\bE|_\cM$ at $\Psi(x)$: 
Neighborhoods of $\Psi(x)$ in $\cM$ are mapped under the restricted linear 
map $\bE|_\cM$ to neighborhoods of $x$ in $\cC$. Figure~\ref{fig1} illustrates 
the absence of openness in Example~\ref{exa:Knauf}. Notice that the cone in 
the example is the state space of a real C*-algebra. The cone is no valid 
choice for the inference of probability measures (\ref{eq:continuous-map}) 
which has the probability simplex $\cP$ as its target set and therefore is
continuous.
\par
Our judgement is that a discontinuity of the ME-inference $\Psi_\theta$ has no 
consequences in statistical physics because it is always marginal. For the 
same reason we think it does not debase $\Psi_\theta$ as a method of quantum 
estimation. More detailed, the {\it relative interior} $\ri(C)$ of a subset $C$ 
of a finite-dimensional real normed vector space is the interior of $C$ in the 
topology of the affine hull of $C$. We denote by $\overline{C}$ the norm 
closure and by $\rb(C):=\overline{C}\setminus\ri(C)$ the {\it relative boundary}. 
Theorem~\ref{thm:openness-condition} and Corollary~\ref{cor:ri_projection} show 
that the inference $\Psi$ is continuous on $\ri(\cC)$. Qualifying the statement 
of marginality, we bring in that discontinuities of $\Psi$ prevent the continuous 
extension of $\Psi|_{\ri(\cC)}$ and are computable simply by taking the norm 
closure of $\Psi(\ri\,\cC)$, see the discussion after 
Lemma~\ref{lem:Gibbs_closure}. Therefore a discontinuity of $\Psi$ causes in its 
neighborhood a very steep functional dependence of $\Psi|_{\ri(\cC)}$ and it is 
not a pure boundary phenomenon. To improve this altogether weak judgement we have 
thought about analyzing the ME-inference of sample averages of iid-states 
\cite{Weis-Knauf}. 
%
%
%
%
\section{Overview}
\label{sec:overview}
\begin{table}
\entrymodifiers={++[F-,]}
\centerline{\xymatrix@R=5mm{%
 *\txt{} & *\txt{} & *\txt{} & *\txt{b)} & 
 \parbox{2.4cm}{\centering exponential\\families} \ar@{-}[dd] \\
 *\txt{\hspace{2ex}a)} & \parbox{2cm}{\centering convex\\geometry} 
 \ar@{-}[r] \ar@{--}@[gray][d] 
 & \parbox{2cm}{\centering openness\\of $\bE|_\cM$} 
 \ar@{-}[r] \ar@{--}@[gray][dl]
 & \parbox{2cm}{\centering continuity\\of $\Psi$} \ar@1{-}[ur]
 \\
 *\txt{} 
 & *++[F--:gray]\txt{\parbox{2cm}{\centering\textcolor{gray}{algebraic\\geometry}}} 
 & *\txt{} & *\txt{} & \parbox{2.4cm}{\centering information\\topology} \ar@{-}[ul] 
\save "2,1"."2,4"*+++[F.:<3ex>]\frm{} \restore
\save "2,4"."1,5"."3,5"*+++[F.:<3ex>]\frm{} \restore
}}
\caption{\label{tab1}Box~a) embraces mainly new matter in 
Section~\ref{sec:inference-continuity}. Box~b) embraces the area of 
ME-inference $\Psi_\theta$ and \rI-topology of an exponential family $\cE$
in Section~\ref{sec:min-rel-inf}: Many earlier results are processed into 
corollaries; a novel generalized \rI-projection is presented in 
Section~\ref{sec:topological-closures}. Methods from a) and b) together raise 
in Section~\ref{sec:unions-of-exponential-families} a question of 
independence of the prior, they discover in Section~\ref{sec:geodesic-closures} 
that the set of inference states $\Psi_\theta(\cC)$ is a geodesic closure of 
$\cE$ and they characterize the norm closure $\overline{\cE}$ in an example in
Section~\ref{sec:upper}.}
\end{table}
\par
This section is an overview of the article and mentions relations to 
other fields. The main idea of the article is the analysis of the
openness of the expected value functional $\bE|_\cM$ which connects 
the subjects in Table~\ref{tab1}.
\par
In the first part, Section~\ref{sec:inference-continuity}, the openness of the 
expected value functional $\bE|_\cM$ is studied in the context of 
Table~\ref{tab1}a). Section~\ref{sec:example} starts with an example of a 
three-dimensional cone \cite{Weis-Knauf}, drawn in Figure~\ref{fig1}. 
Section~\ref{sec:all_disc} proves the equivalence between the continuity of the
inference $\Psi$ and the openness of $\bE|_\cM$ in a more general setting of 
non-linear constraints on a subset of a finite-dimensional real normed vector 
space. A compact and convex subset of such a space will be called {\it convex body}. 
Using convex analysis, we prove in Section~\ref{sec:polytopes} sufficient conditions 
for the openness of a restricted linear map on a convex body.  For example, such a 
map is open if its image is a polytope. This implies for the inference $\Psi$ that
\begin{itemize}
\item[1)]
non-commutative observables $u_1,\ldots,u_k$ are necessary for a 
discontinuous $\Psi$. 
\end{itemize}
See the paragraph of (\ref{eq:support-independence}). We examine the algebras 
${\rm Mat}(2,\bR)\oplus\bR$ and ${\rm Mat}(2,\bC)\oplus\bC$ in 
Section~\ref{sec:solve-cones}. Their state spaces 
are cones where
\begin{itemize}
\item[2)]
discontinuities of $\Psi$ are almost independent of the ranking function
$\phi$.
\end{itemize} 
While 1) shows that discontinuities of $\Psi$ are confined to the quantum 
domain, 2) shows that ranking functions with a discontinuous inference are 
typical. The persistence of the discontinuity in 2) is formulated in 
Remark~\ref{rem:stability} as the question if the continuity of the 
ME-inference $\Psi_\theta$ is independent of the prior
$e^\theta/\tr(e^\theta)$. This holds for arbitrary observables and algebras 
${\rm Mat}(2,\bR)\oplus\bR$ and ${\rm Mat}(2,\bC)\oplus\bC$ as well as for 
the observables of Example~\ref{exa:Knauf} and the algebra 
${\rm Mat}(3,\bC)$. Studying the continuity of $\Psi$ or $\Psi_\theta$ for 
arbitrary observables in arbitrary algebras seems a hard problem. We return
to this question in the last paragraph of this section. 
\par
We mention openness in other fields. A familiar example is the open mapping 
theorem in functional analysis. Papadopoulou \cite{Papadopoulou} calls a convex 
body $K$ {\it stable} if the {\it mid-point map} $K\times K\to K$, 
$(x,y)\mapsto\tfrac 1{2}(x+y)$ is open; we use one of her results in
Remark~\ref{rem:stability}. An analogue notion of stability in a Hausdorff 
locally convex topological vector space is part of the 
{\it Vesterstr\o m-O'Brien theory} in functional analysis, which Protasov and 
Shirokov \cite{Protasov} have extended to spaces such as density matrices on a 
separable Hilbert space. This theory has applications in the analysis of 
entanglement monotones arising from the well-known convex roof construction
which is described for example in~\cite{Uhlmann09}. 
\par
We consider the second part, Section~\ref{sec:min-rel-inf}, in a thoroughness 
due to the particular importance of the ME-inference $\Psi_\theta$ and its 
special context in Table~\ref{tab1}b). The {\it exponential family} 
$\cE:=\{R(\gamma)\mid\gamma\in\bR^k\}$ of canonical states 
$R(\gamma):=e^{\theta+\gamma_1u_1+\cdots+\gamma_ku_k-\Lambda(\gamma)}$ 
for $\gamma=(\gamma_1,\ldots,\gamma_k)\in\bR^k$, where $\Lambda:\bR^k\to\bR$ 
normalizes, is included in the set of ME-inference states $\Psi_\theta(\cC)$
analogous to the case of probability measures (\ref{eq:real-analytic}). It 
seems less well-known that $\bE\circ R:\bR^k\to\ri(\cC)$ is a real analytic 
diffeomorphism onto the interior of $\cC$ if the observables 
$\id,u_1,\ldots,u_k$ are linearly independent. If they are dependent 
then $\cE$ is still a real analytic manifold and $\bE|_{\cE}:\cE\to\ri(\cC)$ 
is a homeomorphism, the {\it mean value chart} of $\cE$. The inverse is a 
real analytic immersion equal to the restricted ME-inference 
\begin{equation}\label{eq:mean-value-par}
\Psi_\theta|_{\ri(\cC)}=\bE|_{\cE}^{-1}:\ri(\cC)\to\cE.
\end{equation}
We prove these statements in Section~6.1 in \cite{Weis-topology}. For the 
maximum-entropy inference $\Psi_0$ we have $\id/\tr(\id)\in\cE$, the 
manifold $\cE$ is known as {\it Gibbsian family} \cite{Petz08} and the 
proofs go back to Wichmann~\cite{Wichmann}. 
\par
Embarking on geodesics and Pythagorean identities one should know that the 
exponential family $\cE$ is an example of Amari's dually flat information geometry 
\cite{Amari}. This branch of differential geometry was investigated in the quantum 
setting by Petz, Nagaoka, Hasegawa, Jen\v cov\'a 
\cite{Petz94,Nagaoka,Hasegawa,Jencova} and others and it has applications such as 
Cram\'er-Rao like inequalities in parameter estimation \cite{Amari-Nagaoka,Petz08}. 
As far as we know, the map $\Psi_\theta$ has not been studied within differential 
geometry. However, the dually flat geometry of $\cE$ provides useful objects. In 
unparametrized form, a {\it $(+1)$-geodesic} is a one-dimensional 
exponential family included in $\cE$ and a {\it $(-1)$-geodesic} is the image of a 
segment in $\ri(\cC)$ under $\Psi_\theta$, see Amari and Nagaoka \cite{Amari-Nagaoka}, 
Section~7.2, and our discussion in \cite{Weis-Knauf}, Section~II. To each geodesic 
one can add two limit points. The union of $(\pm1)$-geodesics in the exponential 
family $\cE$ with their limit points is the {\it $(\pm1)$-geodesic closure} of $\cE$. 
We will use the Pythagorean identity and projection theorem of 
information geometry in their extended forms \cite{Weis-topology}. 
\begin{table}
\begin{tabular}{|c|c|c|c|}
 \hline
 & construction method & more details & non-commutative \\ \hline\hline
 (\ref{eq:inference-ext}) & 
 \multirow{2}{*}{\parbox{4cm}{\centering union of\\exponential families}}
 & extension $\ext$ & yes \\
 (\ref{eq:inv-temp}) & & inverse temperatures & yes \\ \hline
 (\ref{eq:+1-closure}) & \multirow{2}{*}{geodesic closure of $\cE$} 
 & $(+1)$-geodesics & no \\
 (\ref{eq:-1-closure}) & & $(-1)$-geodesics & yes \\ \hline
 (\ref{eq:inference-closure}) 
 & \multirow{4}{*}{\parbox{4cm}{\centering topological closure\\of $\cE$}} 
 & \rI-closure & yes \\
 (\ref{eq:inference-topology}) & & closure in the \rI-topology & yes \\
 (\ref{eq:inference-norm-closure}) & & closure in the norm topology & no \\
 (\ref{eq:generalized-rI}) & & generalized \rI-projection & yes \\ \hline
\end{tabular}
\caption{\label{tab2}These descriptions of the set of ME-inference states 
$\Psi_\theta(\cC)$, relative to an exponential family 
$\cE\subset\Psi_\theta(\cC)$, are discussed in Section~\ref{sec:overview}.
A {\it yes} in the last column means that a description is correct for all 
observables and a {\it no} means it is valid for all commutative but not 
for all non-commutative observables.}
\end{table}
\par
Equation (\ref{eq:mean-value-par}) makes no assertion about the values of the 
ME-inference $\Psi_\theta$ on the relative boundary $\rb(\cC)$. The answers
in Table~\ref{tab2} are in a non-chronological order: Wichmann \cite{Wichmann} 
has recognized the norm closure (\ref{eq:inference-norm-closure}) as a 
super-set of $\Psi_0(\cC)$. Non-commutative discrepancies between the set of 
maximum-entropy states $\Psi_0(\cC)$ and (\ref{eq:+1-closure}) resp.\ 
(\ref{eq:inference-norm-closure}) have stimulated my PhD thesis \cite{Weis-Diss} 
and are also documented in \cite{Weis-Knauf}. For all commutative observables
(\ref{eq:+1-closure}) and (\ref{eq:inference-norm-closure}) are correct 
descriptions of $\Psi_\theta(\cC)$ no matter if the algebra is commutative or 
non-commutative. This is proved in the corresponding sections and follows from
\begin{equation}\label{eq:support-independence}
\cC=\bE(\cM_\cA)=\bE(\cM_\cB),
\end{equation}
which holds for all C*-algebras $\cB\subset\cA$ such that $\cB$ contains the 
identity $\id$ of $\cA$ and the observables $u_1,\ldots,u_k$, see Section~3.4 in 
\cite{Weis-support}. In particular if the observables are commutative, then the 
state space of the C*-algebra generated by $u_1,\ldots,u_k,\id$ is a simplex and 
$\cC=\bE(\cM_\cA)$ is a polytope. A more specialized algebra independence than 
(\ref{eq:support-independence}) holds for the ME-inference: If the prior 
$e^\theta/\tr(e^\theta)$ lies in $\cB$ in addition to $u_1,\ldots,u_k,\id$ then 
$\Psi_\theta(\cC)\subset\cM_\cB$ holds for the ME-inference 
$\Psi_\theta:\cC\to\cM_\cA$. 
\par
We list the content of Section~\ref{sec:min-rel-inf}. 
Section~\ref{sec:progress-convex} starts with a set of orthogonal projections, 
later shown equal to the support projections of ME-inference states. 
Section~\ref{sec:unions-of-exponential-families} defines the extension $\ext$ 
as a union of exponential families, one for each support projection. 
The Pythagorean identity (\ref{eq:Pythagoras}) which we have extended 
in \cite{Weis-topology} from the manifold $\ri(\cM)$ to the convex body $\cM$,
yields $\ext=\Psi_\theta(\cC)$ which is (\ref{eq:inference-ext}). A coordinate 
system of inverse temperatures (\ref{eq:inv-temp}) suits the union $\ext$.
Section~\ref{sec:unions-of-exponential-families} is also the place where 
we scrutinize the Pythagorean identity (\ref{eq:Pythagoras}) very carefully. 
This leads to the question if the continuity of the ME-inference 
$\Psi_\theta$ is independent of the prior $e^\theta/\tr(e^\theta)$.
Section~\ref{sec:geodesic-closures} considers the $(\pm1)$-geodesics introduced 
above. A highlight is the proof, invoking methods from Table~\ref{tab1}a) and b),
that the $(-1)$-geodesic closure (\ref{eq:-1-closure}) of $\cE$ always equals 
$\Psi_\theta(\cC)$.
\par 
The subject of information topology integrates into Table~\ref{tab1}b) 
{\it via} the extended projection theorem (\ref{eq:projection-thm}) in 
Section~\ref{sec:topological-closures}. The projection theorem leads to 
the equality (\ref{eq:inference-closure}) of $\Psi_\theta(\cC)=\cl(\cE)$ 
for the {\it rI-closure} of a subset $X\subset\cM$
\begin{equation}\label{def:cl-rI}
\cl(X):=\{\rho\in\cM\mid\inf_{\sigma\in X}S(\rho,\sigma)=0\}.
\end{equation}
The acronyms of \I{} and \rI{} abbreviate {\it information} and 
{\it reverse information}, respectively, by convention in probability theory 
\cite{CM03}. The \rI-closure is the closure of the topology 
(\ref{eq:inference-topology}) generated by balls of the relative entropy. 
The inclusion $\cl(X)\subset\overline{X}$ holds for all subsets $X\subset\cM$ 
by the {\it Pinsker-Csisz\'ar inequality}
\begin{equation}\label{eq:Pinsker-Csiszar}
\|\rho-\sigma\|^2\leq 2 S(\rho,\sigma), \quad \rho,\sigma\in\cM
\end{equation}
where $\|\cdot\|$ is the trace norm. A proof of (\ref{eq:Pinsker-Csiszar}) 
is given for example in the book by Petz \cite{Petz08}. Clearly the equality 
$\cl(\cE)=\overline{\cE}$ with the norm closure 
(\ref{eq:inference-norm-closure}) holds if and only if $\Psi_\theta$ is 
continuous. The description (\ref{eq:generalized-rI}) in terms of the
\rI-convergence, an\-tici\-pated in the next paragraph, follows with little
effort from the projection theorem (\ref{eq:projection-thm}). 
Section~\ref{sec:commutative} mentions related literature from probability 
theory. Section~\ref{sec:upper} investigates the continuity of the relative 
entropy from an exponential family and applies this to a characterization 
of the norm closure of a Gibbsian family.
\par
Closure discrepancies analogous to $\cl(X)\subsetneq\overline{X}$ in
(\ref{eq:Pinsker-Csiszar}) no not exist for finitely supported probability 
measures but they exist for exponential families of Borel probability 
measures on $\bR^d$ as Csisz\'ar and Mat\' u\v s \cite{CM03} have shown. In 
Section~\ref{sec:commutative} we discuss some aspects of \cite{CM03} and of
earlier literature by \v Cencov, Barndorff-Nielsen and
Tops\o e \cite{Cencov,Csiszar75,Barndorff,Topsoe}. Closure discrepancies 
are connected to the convergence with respect to the relative entropy. A 
sequence of states $(\sigma_i)\subset\cM$ {\it I-converges} to $\omega\in\cM$ 
if  $\lim_{i\to\infty}S(\sigma_i,\omega)=0$ holds and $(\sigma_i)$
{\it rI-converges} to $\omega$ if $\lim_{i\to\infty}S(\omega,\sigma_i)=0$ 
holds. The \rI-convergence is important in this article. We remark that 
Shirokov \cite{Shirokov-charI} has generalized some of Harremo\"es' results 
\cite{Harremoes} about the \I-convergence into the non-commutative setting 
with applications in the analysis of the $\chi$-capacity introduced by 
Holevo \cite{Holevo73}. Leung and Smith \cite{Leung-Smith09} have proved 
with different methods that this functional is continuous for finite-level 
output systems.
\par 
The problematic descriptions (\ref{eq:+1-closure}) and 
(\ref{eq:inference-norm-closure}) of $\Psi_\theta(\cC)$ rise new questions. 
We show in Section~\ref{sec:geodesic-closures} that 
{\it non-exposed faces} (defined in the next section) of the convex support
$\cC$ prevent that $\Psi_\theta(\cC)$ equals the $(+1)$-geodesic closure
(\ref{eq:+1-closure}) of an exponential family $\cE$. In this context
we mention that the state space $\cM$ is a {\it spectrahedron}, that is an 
affine section of a cone of positive semi-definite matrices. Non-exposed 
faces of linear images of spectrahedra play a role in the quantum marginal
problem \cite{Chen12} and non-exposed faces of more general semi-algebraic 
sets are of interest in the foundations of semi-definite programming 
\cite{Netzer-Plaumann-Schweighofer10}. On the other hand, in 
Section~\ref{sec:topological-closures} we show that the topological notion 
of {\it openness} of the expected value functional $\bE|_\cM$ governs the 
continuity of $\Psi_\theta$ as well as the equality of $\Psi_\theta(\cC)$ 
to the norm closure (\ref{eq:inference-norm-closure}). We think the two 
problems are intimately connected because non-exposed faces of $\cC$ are 
indicators of a discontinuous maximum-entropy inference $\Psi_0$ in simple 
examples, see Section~I.B in \cite{Weis-Knauf}. We consider algebraic 
geometry, see Table~\ref{tab1}, a reasonable approach to these problems. 
According to Schweighofer et al.\ \cite{Schweighofer} algebraic and 
algorithmic solutions can be expected to many questions about spectrahedra. 
The openness problem of $\bE|_\cM$ generalizes to the question at which 
points a linear map restricted to a spectrahedron is open. 
%
%
%
%
\section{Real C*-Algebras and Cones}
\label{sec:real-algebras}
\par
We use real algebras because they allow simple examples. The definitions 
of a {\it real C*-algebra} and a {\it real von Neumann algebra} differ from 
the corresponding notions of complex algebras, see for example Alfsen and 
Shultz \cite{Alfsen-Shultz}, only in the field of scalars. References to 
real *-algebras include Ayupov, Rakhimov and Usmanov \cite{Ayupov97} or 
Li \cite{Li03}.
\par
The cone $\cM^{\rm Cone}$ defined in (\ref{def:cone-state-space}) is the 
state space of a real C*-algebra. This fact allows algebraic methods 
including functional calculus and spectral projections in a real C*-subalgebra 
$\cA$ of ${\rm Mat}(n,\bC)$, $n\in\bN$. In particular, our earlier results 
\cite{Weis-support,Weis-Knauf,Weis-topology} all hold for these algebras. 
We point out two properties of real *-algebras: Every finite-dimensional 
real C*-algebra can be represented as a direct sum of full matrix algebras 
with real, complex or quaternionic entries, see for example Theorem 5.7.1 in 
\cite{Li03}. Analogous to normal states and density matrices in a (complex) 
von Neumann algebra, a duality exists between normal states and density 
matrices in a real von Neumann algebra. We comment on this after an example.
\par
We denote the  identity in $\cA$ by $\id$ and we write $\id_n$ and $0_n$ for 
the identity matrix and zero matrix of size $n$, respectively. Denoting the 
{\it Pauli matrices} 
by $\sigma_1:=\left(\begin{smallmatrix}0&1\\1&0\end{smallmatrix}\right)$,
$\sigma_2:=\left(\begin{smallmatrix}0&-{\rm i}\\{\rm i}&0
\end{smallmatrix}\right)$,
$\sigma_3:=\left(\begin{smallmatrix}1&0\\0&-1\end{smallmatrix}\right)$
we write $b\widehat{\sigma}:=b_1\sigma_1+b_2\sigma_2+b_3\sigma_3$ 
for $b=(b_1,b_2,b_3)\in\bR^3$. The real span of $\sigma_1$, $\sigma_2$, 
${\rm i}\sigma_3$ and $\id_2$ defines a real C*-subalgebra $\cA^{\rm Disk}$
of ${\rm Mat}(2,\bC)$. The state space $\cM^{\rm Disk}$ of $\cA^{\rm Disk}$
is the equatorial disk $b_3=0$ in the {\it Bloch ball} \cite{Nielsen-Chuang}
\[
\cM_{{\rm Mat}(2,\bC)}=
\{\tfrac 1{2}(\id_2+b\widehat{\sigma})
\mid b\in\bR^3,
b_1^2+b_2^2+b_3^2\leq 1\}.
\] 
Of course, $\cA^{\rm Disk}\cong{\rm Mat}(2,\bR)$. Another example of a real 
C*-algebra is the direct sum $\cA^{\rm Cone}:=\cA^{\rm Disk}\oplus\bR$ 
embedded into ${\rm Mat}(3,\bC)$ {\it via} block diagonal matrices. The 
four-dimensional real vector space of self-adjoint matrices in 
$\cA^{\rm Cone}$ is spanned by $\sigma_1\oplus 0$, $\sigma_2\oplus 0$, 
$\id_2\oplus 0$ and $\id_3$. The state space of $\cA^{\rm Cone}$ is 
\begin{equation}\label{def:cone-state-space}
\cM^{\rm Cone} :=
\cM_{\cA^{\rm Cone}} =
\conv\left[\{0_2\oplus 1\}\cup(\cM^{\rm Disk}\oplus 0)\right],
\end{equation}
where $\conv$ denotes convex hull. The state space $\cM^{\rm Cone}$ is the 
cone depicted in Figure~\ref{fig1}a). 
\par
A duality exists between normal states $f$ on a real von Neumann algebra $\cB$, 
see for example the characterization in Theorem 4.5.3 in \cite{Li03}, and density 
matrices $\rho$ in $\cB$, that is trace class operators of the form $\rho=a^*a$,
$a\in\cB$, $\tr(\rho)=1$, such that $f(b)=\tr(b\rho)$ holds for all $b\in\cB$. 
Thereby a {\it state} is a linear functional $f:\cB\to\bR$ such that 
$f(a^*a)\geq 0$ holds for all $a\in\cB$ and such that $f(a)=0$ holds for all 
skew-symmetric matrices $(a^*=-a)$. One can pass to the complexification of 
$\cB$, apply the corresponding result in the complex case, and restrict it to 
$\cB$. The linear functional $\cA^{\rm Disk}\to\bR$, 
$a\mapsto\tr(a\,{\rm i}\sigma_3)$ shows that the condition about skew-hermitian 
matrices is necessary. 
\par
Cones like $\cM^{\rm Cone}$ are among the simplest examples where a restricted 
linear map may not be open. Therefore we introduce some notation about cones. 
The {\it closed segment} between two points $x,y$ in a finite-dimensional real 
vector space $X$ is $[x,y]:=\{(1-\lambda)x+\lambda y\mid 0\leq\lambda\leq 1\}$, 
the {\it open segment} is $]x,y[\,:=\{(1-\lambda)x+\lambda y\mid 0<\lambda<1\}$. A 
{\it face} of a convex subset $C\subset X$ is a convex subset $F\subset C$ such 
that every segment $[x,y]\subset C$ with $]x,y[\,\cap F\neq\emptyset$ is 
included in $F$. A one-point face (face of dimension zero) is called 
{\it extremal point} and a face of dimension $\dim(C)-1$ is called {\it facet}. 
A subset $F\subset C$ is an {\it exposed face} of $C$ if $F=\emptyset$ or if $F$ 
equals the set of maximizers in $C$ of a linear functional on $X$. One can show 
that every exposed face is a face. A face which is not exposed is called 
{\it non-exposed face}. 
\par
A {\it cone} is defined as the convex hull of the union of a convex body $B$ 
with a point $a$ not in $\aff(B)$, the affine hull of $B$. The set 
$B$ is the ${\it base}$, the relative boundary of $B$ is the {\it directrix}, 
$a$ is the {\it apex} and each segment $[x,a]$, where $x$ belongs to the 
directrix, is a {\it generatrix} of the cone. If the base $B$ is a 
{\it solid ellipsoid} (affinely isomorphic to a Euclidean unit ball), then all 
faces of the cone are exposed and the set of extremal points consists of the 
elements of the directrix and of the apex. If $\dim(B)\geq 2$ then, apart from 
extremal points, the faces of the cone are $\emptyset$, the generatrices 
(one-dimensional), the base $B$ (facet, $\dim(B)>1$) and the cone itself. 
%
%
%
%
\section{Openness of Restricted Linear Maps}
\label{sec:inference-continuity}
\par
This section explores a local continuity condition of the inference
in terms of the openness of the expected value functional. The openness 
condition works in more general settings of parametrized constraints, 
assuming a unique global maximum on each constraint set. Sufficient 
conditions of the openness and their corollaries are discussed in 
further detail in the overview in Section~\ref{sec:overview}.
\begin{Def}[Optimization under non-linear constraints]
\label{def:basic}
Let $f:V\to W$ be a continuous map between finite-dimensional real normed 
vector spaces $V,W$ and let $K\subset V$ be compact. Then $L:=f(K)\subset W$ 
parametrizes the fibers of $f|_K$ and we define
\[
F:L\to 2^K,\qquad
w\mapsto f|_K^{-1}(w)=\{v\in K\mid f(v)=w\},
\]
where $2^K$ denotes the power set of $K$. Let $g:K\to\bR$ be a continuous real 
valued function, the {\it objective functional}. We define the {\it maximum}
\[
h:L\to\bR,\qquad
w\mapsto\max\{g(v)\mid v\in F(w)\}.
\]
Assuming that $g$ has on each fiber $F(w)$, $w\in L$, a unique maximum, we 
define the {\it maximizer}
\[
H:L\to K,\qquad
w\mapsto\argmax\{g(v)\mid v\in F(w)\}.
\]
\end{Def}
%
%
%
%
\subsection{A Minimal Example}
\label{sec:example}
\par
We explain our minimal example of a discontinuous maximum-entropy inference 
\cite{Weis-Knauf}. We show that openness of $f|_K$ is necessary for the 
continuity of $H$.
\par
To analyze the continuity of the maximum and maximizer, we define in a topological 
space $X$ an {\it open neighborhood} of a point $x\in X$ as an open subset of $X$ 
containing $x$. A {\it neighborhood} of $x$ is a subset of $X$ containing an open 
neighborhood of $x$. We will consider $K$ and $L$ with the subspace topology 
induced by the norm topology on $V$ and $W$, respectively.  For example, if $L$ is
the closed unit disk in $\bR^2$ then $L$ has no {\it boundary} and its 
{\it relative boundary}, defined in the last paragraph of the introduction, is the 
unit circle $S^1$. 
\begin{Def}[Openness]
\label{def:open_map}
The restricted map $f|_K:K\to L$ is {\it open} at $v\in K$ if for any neighborhood 
$N\subset K$ of $v$ the image $f(N)$ is a neighborhood of $f(v)$ in $L$. We say 
$f|_K$ is {\it open on} a subset $X\subset K$ if $f|_K$ is open at each $v\in X$ 
and $f|_K$ is {\it open} if $f|_K$ is open on $K$.
\end{Def}
\par
A necessary continuity condition is immediate.
\begin{Lem}
\label{lem:one_direction}
If the maximizer $H:L\to K$ is continuous at $w\in L$, then the restricted map 
$f|_K:K\to L$ is open at $H(w)$. 
\end{Lem}
{\it Proof:}
Let $w\in L$ and let $N$ be a neighborhood of $H(w)$. If $H$ is continuous at 
$w\in L$ then $H^{-1}(N)$ is a neighborhood of $w$. We have 
$f(N)\supset f(N\cap H(L))=H^{-1}(N)$, so $f(N)$ is a neighborhood of 
$f(H(w))=w$. This proves the claim.
\hspace*{\fill}$\square$\\
\par
The following example is based on a three-dimensional convex cone 
$K=\cM^{\rm Cone}$ defined in Section~\ref{sec:real-algebras} as the state 
space of a real C*-algebra. So Gibbsian families are defined. It is a minimal 
example of a discontinuous maximizer $H$ under linear constraints $f$, 
because $H$ is continuous for $\dim(K)\leq 2$, for polytopes $K$ and for 
Euclidean balls $K$ by Theorem~\ref{thm:openness-condition} applied to 
Example~\ref{exa:strong-sufficient-examples}.1, to 
Corollary~\ref{cor:open_poly} and to 
Example~\ref{exa:strong-sufficient-examples}.2, respectively. 
\begin{Exa}[Discontinuous maximum-entropy inference]
\label{exa:Knauf}
Figure~\ref{fig1}a) shows the cone $\cM^{\rm Cone}$, defined in 
(\ref{def:cone-state-space}). The apex is $0_2\oplus 1$ and the directrix 
is the circle para\-met\-rized by 
$\rho(\alpha):=\tfrac 1{2}(\id_2+\sin(\alpha)\sigma_1+\cos(\alpha)\sigma_2)
\oplus 0$, $\alpha\in\bR$. We choose two observables 
\[
u_1:=\sigma_1\oplus 0
\quad\text{and}\quad
u_2:=\sigma_2\oplus 1
\]
and set ${\bf u}:=(u_1,u_2)$. The convex support $\cC=\bE(\cM^{\rm Cone})$
in Definition~\ref{def:entropic-inference} is the unit disk. Since 
$\cM^{\rm Cone}$ is the state space of an algebra it includes the Gibbsian 
family 
\[
\cE:=\{e^\nu/\tr(e^\nu)\mid \nu=\gamma_1u_1+\gamma_2u_2\text{ for }
\gamma_1,\gamma_2\in\bR\},
\] 
called {\it Staffelberg family} in \cite{Weis-Knauf}. The algebraic origin of 
$\cM^{\rm Cone}$ also implies that the image of the maximum-entropy inference 
$\Psi_0$ is included in $\cM^{\rm Cone}$ independent of the choice of one of 
the algebras 
$\cA^{\rm Cone}\subsetneq{\rm Mat}(2,\bC)\oplus\bC\subsetneq{\rm Mat}(3,\bC)$,
as we have discussed in the paragraph of (\ref{eq:support-independence}) in
Section~\ref{sec:overview}. The set of maximum-entropy inference states is 
computed in Theorem~18 and Theorem~21 in \cite{Weis-Knauf} and equals 
\[
\Psi_0(\cC)=\cE\cup\{\rho(\alpha)\mid\alpha\in\,]0,2\pi[\,\}\cup\{c\}
\]
for $c:=\tfrac 1{2}(\rho(0)+0_2\oplus 1)$. The generatrix 
$[\rho(0),0_2\oplus 1]$ of $\cM^{\rm Cone}$ is the fiber of 
$\bE|_{\cM^{\rm Cone}}$ at $\bE(c)=(0,1)$. The states $\rho(0)$ and 
$0_2\oplus 1$ are orthogonal rank-one projections so the von Neumann 
entropy has maximal value $\log(2)$ on this fiber at $\Psi_0(0,1)=c$. 
Since points $x\neq\bE(c)$ on the unit circle have 
inference values $\Psi_0(x)$ on the directrix of $\cM^{\rm Cone}$, 
the maximum-entropy inference $\Psi_0$ is not continuous at $(0,1)$. 
\par
We arrive at the same conclusion from the openness condition in 
Lemma~\ref{lem:one_direction}. We put $u_3:=0_2\oplus 1-\rho(0)$. Then a 
neighborhood of $c$ is defined by
\[
N:=\{\rho\in\cM^{\rm Cone}\mid\langle\rho,u_3\rangle\geq-1/3\}.
\]
Figure~\ref{fig1}b) illustrates that $\bE(N)$ is not a neighborhood of 
$(0,1)$ in $\cC$, so $\bE|_{\cM^{\rm Cone}}$ is not open at $c$ and by
the lemma $\Psi_0$ is not continuous at $(0,1)$. 
\end{Exa}
%
%
%
%
\subsection{The Local Continuity Condition}
\label{sec:all_disc}
\par
We show that the continuity condition from Lemma~\ref{lem:one_direction}
is sufficient. The idea stems from Wichmann's Theorem~2d) in 
\cite{Wichmann} which as a byproduct yields a global continuity condition.
\par
A function $\varphi:X\to[-\infty,+\infty]$ on a metric space $X$ is 
{\it upper semi-continuous} at $x\in X$ if 
$\varphi(x)\geq\limsup_{i\to\infty}\varphi(x_i)$ for every sequence 
$(x_i)\subset X$ such that $x=\lim_{i\to\infty}x_i$ and
$\varphi$ is {\it lower semi-continuous} at $x$ if $-\varphi$ is upper 
semi-continuous at $x$.
\begin{Lem}[Upper semi-continuity of the maximum]~
\label{lem:upper-semi-maximum}
The maximum $h:L\to\bR$ is upper semi-continuous on $L$. For all $w\in L$ the 
maximizer $H:L\to K$ is continuous at $w$ if and only if $h$ is continuous at 
$w$.
\end{Lem}
{\it Proof:}
Two references after this lemma prove the upper semi-continuity of $h$. We 
prove the continuity of $h$ in a larger context such as to recycle it in 
Lemma~\ref{lem:upper-semi-divergence}. Let $\tilde X$ be a metric space, 
$\tilde f:\tilde X\to L$ a continuous function and set
$\tilde\Pi:=H\circ\tilde f$. We show if $h\circ\tilde f$ is continuous at 
$x\in\tilde X$ then $\tilde\Pi$ is continuous at $x$. Let 
$(x_i)_{i\in\bN}\subset\tilde X$ be a sequence with $x=\lim_{i\to\infty}x_i$. 
First we assume that $\tilde\Pi(x_i)$ converges. Then 
$g(\lim_{i\to\infty}\tilde\Pi(x_i))=g(\tilde\Pi(x))$ and
$f(\lim_{i\to\infty}\tilde\Pi(x_i))=f(\tilde\Pi(x))$ prove 
$\tilde\Pi(x_i)\rightarrow\tilde\Pi(x)$ because $g$ has a unique 
maximizer in each fiber of $f$. Since $K$ is compact the assumed convergence 
of $\tilde\Pi(x_i)$ is no restriction and the claim follows.
The choices $\tilde X:=L$ and $\tilde f:={\rm Id}|_L$ suffice to show that 
the continuity of $h$ at $w\in L$ implies the continuity of $H$ at $w$.
The equation $h=g\circ H$ proves the converse as $g$ is continuous.
\hspace*{\fill}$\square$\\
\par
Compactness of $K$ is essential for the upper semi-continuity of $h$ in 
Lemma~\ref{lem:upper-semi-maximum}, as the example 
$K:=\{(0,0)\}\cup\{(x,y(x))\mid 0<x\leq 1\}\subset\bR^2$ for $y(x):=1/x$ or 
$y(x):=1-x$ and $f(x,y):=x$, $g(x,y):=y/(y+1)$ shows. The upper semi-continuity 
of $h$ is proved in the first lines of Proposition~4 in \cite{Lima} using closed 
level sets. The proof in Theorem~2 in Section VI.3 in \cite{Berge} uses open 
coverings and generalizes easily to $[-\infty,+\infty]$-valued functions 
(for a constant constraint set $K_2$): 
\begin{Rem}
\label{rem:compact-infimum-lower}
Let $K_1,K_2\subset X$ for a finite-dimensional real normed vector space $X$ and 
let $K_2\neq\emptyset$ be compact. If $\varphi:K_1\times K_2\to[-\infty,+\infty]$ 
is a lower semi-continuous function then the minimum $K_1\to[-\infty,+\infty]$,
$v\mapsto\min\{\varphi(v,w)\mid w\in K_2\}$ is a lower semi-continuous function.
This statement will be used in Theorem~\ref{thm:d-continuity}. 
\end{Rem}
\par
We come back to the continuity condition.
\begin{Lem}
\label{lem:Wichmann}
Let $(w_i)\subset L$ be a sequence converging to $w\in L$. If there is
a sequence $(v_i)\subset K$ such that $v_i\in F(w_i)$ for all $i\in\bN$ and 
such that $v_i\stackrel{i\to\infty}{\rightarrow}H(w)$, then 
$H(w_i)\stackrel{i\to\infty}{\rightarrow}H(w)$.
\end{Lem}
{\it Proof:}
The continuity of $g$ implies $g(v_i)\rightarrow g\circ H(w)=h(w)$. Since for 
all $w'\in L$ the number $h(w')$ maximizes $g(v')$ among all $v'\in F(w')$ and 
by the upper semi-continuity of $h$, see Lemma~\ref{lem:upper-semi-maximum}, 
we have
\[
h(w)=\lim_{i\to\infty}g(v_i)
\leq \lim\inf_{i\to\infty}h(w_i)
\leq \lim\sup_{i\to\infty}h(w_i)
\leq h(w).
\]
This proves $h(w_i)\rightarrow h(w)$ and $H(w_i)\rightarrow H(w)$ follows 
from Lemma~\ref{lem:upper-semi-maximum}.
\hspace*{\fill}$\square$\\
\par
A generalization of Wichmann's Theorem 2d) in \cite{Wichmann} now follows 
as a corollary.
\begin{Lem}
\label{lem:Gibbs_closure}
If $K$ is a convex body and if $f$ is linear,
then $H(L)\subset\overline{H({\rm ri}(L))}$.
\end{Lem}
{\it Proof:} Let $w\in L$ and let $(v_i)\subset{\rm ri}(K)$ be a sequence 
converging to $H(w)\in K$. The points $w_i:=f(v_i)$ all lie in ${\rm ri}(L)$ 
for $i\in\bN$, as $f({\rm ri}(K))={\rm ri}(f(K))={\rm ri}(L)$, see for
example Theorem 6.6 in \cite{Rockafellar}. Lemma~\ref{lem:Wichmann} completes 
the proof.
\hspace*{\fill}$\square$\\
\par
Lemma~\ref{lem:Gibbs_closure} shows that a discontinuity of $H$ can not be
removed by changing values of $H$ on the relative boundary $\rb(L)$. The 
lemma also implies a global continuity condition of $H$: Since $L$ is compact, 
$H$ is continuous if and only if $\overline{H(L)}=H(L)$. The lemma implies 
$\overline{H({\rm ri}\,L)}=\overline{H(L)}$, so $H$ is continuous if and only 
if $\overline{H({\rm ri}\,L)}=H(L)$. In the case of the ME-inference 
$\Psi_\theta$ this condition is, according to (\ref{eq:mean-value-par}), the
equality between the norm closure of an exponential family $\cE$ and the set 
of ME-inference states,  $\overline{\cE}=\Psi_\theta(\cC)$. We prove the 
openness condition. 
\begin{Thm}
\label{thm:openness-condition}
For any $w\in L$ the maximizer $H$ is continuous at $w$ if and only if the
restricted map $f|_K$ is open at $H(w)$. 
\end{Thm}
{\it Proof:}
As a metric space is first countable, there exists a local case 
$\{B_i\}_{i\in\bN}$ of the topology at $H(w)\in K$. Let us assume a sequence 
$(w_i)\subset L$ converges to $w$. By assumptions, $f|_K$ is open at $H(w)$ so 
there exists for each $j\in\bN$ a natural number $n_j\in\bN$ such that for all 
$i\geq n_j$ we have $w_i\in f(B_j)$. Without loss of generality we can assume 
that $\{B_i\}$ is monotonically decreasing and $(n_j)$ is strictly 
monotonically increasing. Then for every $i\in\bN$ exists a unique $j\in\bN$ 
such that $n_j\leq i<n_{j+1}$ and we select an arbitrary $v_i\in B_j\cap F(w_i)$.
Now $v_i\rightarrow H(w)$ and $H(w_i)\rightarrow H(w)$ follows from
Lemma~\ref{lem:Wichmann}. The converse is proved in 
Lemma~\ref{lem:one_direction}.
\hspace*{\fill}$\square$\\
%
%
%
%
\subsection{Sufficient Conditions for Openness}
\label{sec:polytopes}
\par
In this section we prove sufficient conditions for the openness of 
the restricted map $f|_K$ from Definition~\ref{def:basic}, provided 
$K\subset V$ is a convex body and $f:V\to W$ is linear. 
\begin{Def}
Let $C\subset X$ be a convex subset of a finite-dimensional real normed
vector space $X$. The {\it gauge} of $C$ is defined by
\[
\gamma_C(v):=\inf\{\lambda\geq 0\mid v\in\lambda C\},
\quad v\in X.
\]
A function $\gamma:X\to\bR\cup\{+\infty\}$ is {\it positively homogeneous} 
if for all $\lambda>0$ and $u\in X$ we have 
$\gamma(\lambda u)=\lambda\gamma(u)$. Given a non-empty subset $Y\subset X$, 
the {\it positive hull} of $Y$ is defined by 
${\rm pos}(Y):=\{\lambda v\mid\lambda\geq 0, v\in Y\}$. We denote the norm 
in any normed vector space by $\|\cdot\|$. For a point $y\in Y$ and a 
positive real $r>0$, we define the {\it closed ball} respectively 
{\it sphere}
\[
B_{r}(y,Y):=
\{x\in Y\mid\|x-y\|\leq r\}
\quad\text{resp.}\quad
S_r(y,Y):=
\{x\in Y\mid\|x-y\|=r\}.
\]
\end{Def}
\par
We recall that gauge generalizes norm in the sense that the gauge of the 
unit ball in $X$ equals the norm. If $C\subset X$ is convex 
then $\gamma_C$ is convex and positively homogeneous. If $u$ is normalized, 
then $\gamma_{C-x}(u)$ is the inverse radius of $C$ from the center $x\in C$ 
in the direction $u$, see for example Rockafellar~\cite{Rockafellar}. 
\begin{Pro}
\label{pro:gauge_condition}
Let $K\subset V$ be a convex body, let $f:V\to W$ be a linear map and
let $w\in L=f(K)$. If the gauge $\gamma_{L-w}$ is bounded on the unit 
sphere $S_1(0,{\rm pos}(L-w))$, then $f|_K$ is open on the fiber $F(w)$.
\end{Pro}
{\it Proof:}
Because of the bounded gauge, a sufficiently small ball about $w$ is 
a union of uniform length segments: There exists $\epsilon>0$ such that
\[
B_{\epsilon'}(w,L)=\bigcup\{[w,w']\mid w'\in S_{\epsilon'}(w,L)\},
\quad
0<\epsilon'\leq\epsilon.
\]
For every $v\in F(w)$ in the fiber $F(w)\subset K$ and diameter 
$\delta:=\max_{v',v''\in K}\|v'-v''\|$ we have
\[
f(B_\delta(v,K))=f(K)=L
\supset B_\epsilon(w,L).
\]
Since $B_\epsilon(w,L)$ is a union of segments as above, and since $K$
is convex, the inclusion $f(B_\delta(v,K))\supset B_\epsilon(w,L)$ holds
after scaling both $\delta$ and $\epsilon$ with the same number in 
$[0,1]$. For arbitrary $\delta'>0$ with $\delta'\leq\delta$ the number 
$\epsilon':=\epsilon\cdot\delta'/\delta$ is strictly positive. Scaling 
by $\delta'/\delta$ shows $f(B_{\delta'}(v,K))\supset B_{\epsilon'}(w,L)$ 
so $f(B_{\delta'}(v,K))$ is a neighborhood of $w$. This proves that 
$f|_K$ is open at $v$.
\hspace*{\fill}$\square$\\
\par
Proposition~\ref{pro:gauge_condition} is not a necessary condition for the 
openness of $f|_K$ as the Example~\ref{exa:strong-sufficient-examples}.2 of a 
ball shows. More examples can be constructed from an arbitrary convex 
body $L$, the cylinder $K:=L\times[0,1]$ and $f(w,\lambda):=w$, $w\in L$, 
$\lambda\in[0,1]$. Another example is the apex of a cone.
\begin{Cor}
\label{cor:ri_projection}
Let $K\subset V$ be a convex body and let $f:V\to W$ be a linear map.
Then $f|_K$ is open on the fiber $F(w)$ for all relative interior points 
$w\in{\rm ri}(L)$.
\end{Cor}
{\it Proof:} 
If $w\in\ri(L)$ then $L$ contains the closed ball $B_\epsilon(w,{\rm aff}(L))$
for some $\epsilon>0$, where ${\rm aff}(L)$ denotes the affine hull of $L$. Hence 
$\gamma_{L-w}$ is bounded on the unit sphere $S_1(0,{\rm pos}(L-w))$ by 
$1/\epsilon$ and Proposition~\ref{pro:gauge_condition} proves the claim.
\hspace*{\fill}$\square$\\
\par
We recall that a {\it closed half-space} in a finite-dimensional real vector 
space $X$ is defined by $\{x\in X\mid l(x)\leq \beta\}$, where $l:X\to\bR$ is 
a non-zero linear functional and $\beta\in\bR$. A {\it polyhedral convex set} 
in $X$ is a finite intersection of closed half-spaces and a {\it polytope} is 
a bounded polyhedral convex set. It is well-known, see for example Remark 3.1 
in \cite{Voigt}, that a polyhedral convex subset $C\subset X$ containing the 
origin has bounded gauge $\gamma_C$ on the unit sphere $S_1(0,{\rm pos}(C))$.
\begin{Cor}
\label{cor:open_poly}
Let $K\subset V$ be a convex body and let $f:V\to W$ be a linear map.
If $L=f(K)$ is a polytope, then the restricted linear map $f|_K$ is open.
\end{Cor}
{\it Proof:}
As mentioned in the previous paragraph, for all $w\in L$ the gauge 
$\gamma_{L-w}$ is bounded on $S_1(0,{\rm pos}(L-w))$. 
Proposition~\ref{pro:gauge_condition} completes the proof.
\hspace*{\fill}$\square$\\
\par 
Another openness condition is injectivity. 
\begin{Lem}
\label{lem:one-point-fibers}
Let $K\subset V$ be a convex body and let $f:V\to W$ be a linear map.
Let $v\in K$ lie in a one-point fiber, that is $\{v\}=F(f(v))$ holds. 
Then the restricted linear map $f|_K$ is open at $v$.
\end{Lem}
{\it Proof:} 
We show continuity of $H$ at $w:=f(v)$ then Theorem~\ref{thm:openness-condition} 
will complete the proof. We define the objective functional $g$ as the 
restriction to $K$ of the quadratic form $V\to\bR$, $v'\mapsto-\|v'\|^2$, 
associated to a Euclidean norm $\|\cdot\|$ on $V$. The quadratic form is 
continuous and strictly concave so $g$ satisfies the assumptions of the 
theorem. Since $K$ is closed and the fiber $F(x)=\{v\}$ contains a unique
point, the maximizers $H(w_i)$ of any converging sequence $(w_i)\subset L$ 
with limit $w$ must converge to $H(w)=v$. This proves the continuity of 
$H$ at $w$.
\hspace*{\fill}$\square$\\
\par
Further examples demonstrate the results in this section.
\begin{Exa}[Openness of a linear map $f$ restricted to a convex body $K$]~
\label{exa:strong-sufficient-examples}
\begin{enumerate}
\item
If $\dim(K)\leq 2$, then $f|_K$ is open. Indeed, if 
$\dim(L)\leq 1$, then $L$ is a polytope and openness follows. 
Otherwise $f|_K$ is injective.
\item 
If $K$ is a solid ellipsoid, then $f|_K$ is open. It is sufficient to prove 
this for the full-dimensional unit ball centered at the origin and for an
orthogonal projection $f$. Every relative boundary point $w\in\rb(L)$ is
normalized so $F(w)=\{w\}$ is a one-point fiber.
\item
Let $G$ be a facet of $L$, that is a face of dimension $\dim(L)-1$.
If $w$ is a relative interior point of $G$, then $f|_K$ is open on $F(w)$ 
because the gauge of $L-w$ is bounded on the unit sphere 
$S_1(0,{\rm pos}(L-w))$.
\end{enumerate}
\end{Exa}
%
%
%
%
\subsection{Cones and Typicality of Discontinuities}
\label{sec:solve-cones}
\par
We study the openness of a linear function restricted to a cone, using the 
terminology of cones and faces from Section~\ref{sec:real-algebras}. This 
allows a continuity analysis of the inference $\Psi$ in 
${\rm Mat}(2,\bR)\oplus\bR\cong\cA^{\rm Cone}$ and 
${\rm Mat}(2,\bC)\oplus\bC$. We conclude that ranking functions with a 
discontinuous inference are typical. 
\begin{Lem}\label{lem:cones-non-open}
Let $B$ be a solid ellipsoid, $a\not\in\aff(B)$ and consider the cone
$K:=\conv(B,a)\subset W$ with base $B$ and apex $a$. If $w\in L$ and 
if the restricted linear map $f|_K$ is not open on $F(w)$ then $F(w)$ 
is a generatrix of $K$.
\end{Lem}
{\it Proof:} 
There exists a unique face $G$ of $L$ such that $w\in\ri(L)$ and the 
inverse image $\widetilde{G}:=F(G)$ is a face of $K$, see the paragraph 
of (\ref{def:face-projection}). If $f|_K$ is not open on the fiber 
$F(w)$ then $f|_{F(w)}$ is not injective by 
Lemma~\ref{lem:one-point-fibers}. Neither is $f|_{\widetilde{G}}$ so 
$\dim(\widetilde{G})>\dim(G)\geq 0$ holds. If $\dim(\widetilde{G})>1$, 
then by the discussion of faces of $K$ in 
Section~\ref{sec:real-algebras}, we have $\widetilde{G}=K$ or
$\widetilde{G}=B$. 
\par
If $\widetilde{G}=K$ then $G=L$. If $\widetilde{G}=B$ then $G=f(B)$
and $L=\conv(G,f(a))$. Assuming $f(a)\in\aff(G)$ we have $f(a)\in G$ 
because $G$ is a face of $L$. In this case $G=L$. Otherwise if 
$f(a)\notin\aff(G)$ holds then $G$ is a facet of $L$. In all cases 
$w$ is a relative interior point of $L$ or of a facet of $L$. In either 
case Corollary~\ref{cor:ri_projection} or 
Example~\ref{exa:strong-sufficient-examples}.3 gives the contradiction 
that $f|_K$ is open on $F(w)$. So $\dim(\widetilde{G})=1$ and $\dim(G)=0$
hold. As Example~\ref{exa:strong-sufficient-examples}.1 shows 
$\dim(K)\geq 3$, the discussion of faces of $K$ in 
Section~\ref{sec:real-algebras} shows that $\widetilde{G}$ is a generatrix.
\hspace*{\fill}$\square$\\
\par
The proof of the preceding lemma works for a strictly convex compact set
in place of a solid ellipsoid. This is not clear in the next lemma.
\begin{Lem}
\label{lem:openness-cone-projection}
Let $B$ be a solid ellipsoid, $a\not\in\aff(B)$ and consider the cone
$K:=\conv(B,a)\subset W$ with base $B$ and apex $a$. The restricted linear 
map $f|_K$ is not open if and only if $\dim(L)\geq 2$ and a generatrix
of $K$ is a fiber of $f|_K$, that is $[v,a]=F\circ f(a)$ holds for 
some $v\in\rb(B)$. In that case $f|_K$ is not open at all points of 
$\,]v,a]:=\{(1-\lambda)v+\lambda a\mid\lambda\in\,]0,1]\}$ and open on
the complement $K\setminus\,]v,a]$.
\end{Lem}
{\it Proof:} 
Notice from Corollary~\ref{cor:open_poly} that $\dim(L)\geq 2$ is necessary
if $f|_K$ is not open since $L$ is a polytope otherwise. By 
Lemma~\ref{lem:cones-non-open} it is also necessary that $F(w)$ is a 
generatrix for some $w\in L$. We prove the converse by exhibiting the
claimed points of non-openness and openness. 
\par
Since openness is equivariant under invertible linear maps by
Lemma~\ref{lem:one-point-fibers} we can transform $K$ into $\bR^{d+1}$ such 
that the base $B$ of $K$ is the $d$-dimensional Euclidean unit ball about 
$0$ in the hyperplane 
$H:=\{(x_1,\ldots,x_{d+1})\in\bR^{d+1}\mid x_{d+1}=0\}$, the above generatrix 
is $F(w):=[v,a]$ for $v:=(1,0,\ldots,0)$ and $a:=(1,0,\ldots,0,1)$. Preserving
the kernel of $f$ we can also assume that $f$ is the orthogonal projection 
onto a subspace $U\subset H$. Corollary~\ref{cor:ri_projection} shows that 
$f(v)$ is a relative boundary point of the unit ball $L=f(K)$. So we have 
$v\in U$ as well as $w=f(v)=v$.
\par
Since generatrices intersect in the apex $a$, Lemma~\ref{lem:cones-non-open}
proves that $f|_K$ is open on $F(L\setminus\{w\})$. We study the points $b$ 
on the fiber $F(w)=[v,a]$ using Theorem~\ref{thm:openness-condition}. We use
objective functional $g(v):=-\|v-b\|^2$ for a Euclidean norm $\|\cdot\|$.
For all $w'\neq w$ in the relative boundary $\rb(L)$ the maximizer is 
$H(w')=w'\in U$ while $H(w)=b$ holds. If $b\in[v,a]$ and $b\neq v$ then 
$b\not\in U$ and given that $\dim(U)\geq 2$ holds the function $H$ is 
discontinuous at $w$. Then Theorem~\ref{thm:openness-condition} shows that 
$f|_K$ is not open at $b$. Clearly $f|_K$ is open at $v$ because $v$ is in 
the base of the cone and $f(v)=f(a)$ holds. This completes the proof.
\hspace*{\fill}$\square$\\
\par
In the discussion whether a discontinuous inference is typical we
consider also complex C*-algebras as they are models of quantum systems 
in theoretical physics. In Section~\ref{sec:progress-convex} 
we continue this analysis in ${\rm Mat}(3,\bC)$.
\begin{Exa}\label{exa:typical1}
Cone state spaces demonstrate that for certain choices of observables the 
inference $\Psi$ is discontinuous for almost all ranking functions.
For the direct sum algebras $\cA=\cA^{\rm Cone}$, considered in 
Example~\ref{exa:Knauf}, $\cA={\rm Mat}(2,\bR)\oplus\bR\cong\cA^{\rm Cone}$
and $\cA={\rm Mat}(2,\bC)\oplus\bC$ the state space $\cM=\cM_\cA$ is a 
symmetric cone with base a Euclidean ball. Theorem~\ref{thm:openness-condition} 
and Lemma~\ref{lem:openness-cone-projection} imply for these examples: The 
inference $\Psi:\cC\to\cM$ is not continuous, if and only if $\Psi$ is 
discontinuous at a single point $x\in\cC$. Then the linear family
$\cL_x=\bE|_\cM^{-1}(x)=[\rho,\sigma]$ is a generatrix of the cone $\cM$
for $\rho=\rho_0\oplus 0$ and $\sigma=:0_2\oplus 1$ where $\rho_0$ is a state
in ${\rm Mat}(2,\bC)$. We have seen $\Psi_0$ is discontinuous in 
Example~\ref{exa:Knauf} at $(0,1)$ with linear family
$\cL_{(0,1)}=[\rho(0),\sigma]$.
\par
Moreover, if we fix observables where $\Psi$ is discontinuous at $x\in\cC$
and where $\bE|_\cM^{-1}(x)=[\rho,\sigma]$, then for all ranking functions $\phi$
the inference $\Psi$ is discontinuous at $x$ if and only if $\Psi(x)\neq\rho$
while $\Psi$ is continuous on $\cC\setminus\{x\}$. In the sense that the equality 
$\Psi(x)=\rho$ is exceptional, we say that $\Psi$ is discontinuous for
{\it almost all} ranking functions. 
\end{Exa}
%
%
%
%
\section{Maximum-Entropy States}
\label{sec:min-rel-inf}
\par
This section collects descriptions of the set of ME-inference states 
$\Psi_\theta(\cC)$ in a non-commutative algebra. A detailed overview is 
given in Section~\ref{sec:overview}.
%
%
%
%
\subsection{Support Projections}
\label{sec:progress-convex}
\par
We recall our geometric construction \cite{Weis-support} of a set of 
orthogonal projections $\cP({\bf u})$ in the algebra $\cA$ from the 
observables ${\bf u}=(u_1,\ldots,u_k)$. As an example we discuss 
typicality of a discontinuous inference in the algebra ${\rm Mat}(3,\bC)$.
\par
We will see in Section~\ref{sec:unions-of-exponential-families} that 
$\cP({\bf u})$ is the set of support projections of all ME-inference states. 
The support projection of a self-adjoint matrix $a\in\sa$ is the orthogonal 
projection ($p=p^2=p^*$) in $\cA$ which is the sum of all spectral projections 
of $a$ corresponding to non-zero eigenvalues. The geometric construction of
$\cP({\bf u})$ uses lattice morphisms: One is defined by $\bE|_\cM^{-1}$ and
maps faces of the convex support $\cC=\bE(\cM)$ to faces of the state space 
$\cM=\cM_\cA$, partially ordered by inclusion. The second map is the 
isomorphism between the faces of the state space $\cM$ and the projections 
in the algebra $\cA$, see Alfsen and Shultz \cite{Alfsen-Shultz}.
\par
The geometric construction of $\cP({\bf u})$ is needed explicitly in 
Lemma~\ref{lem:I-projection-range}. The family of relative interiors of faces 
of the convex support $\cC$ is a partition of $\cC$, see Theorem 18.2 in 
\cite{Rockafellar}. So for $x\in\cC$ there is a unique face $\tilde{F}(x)$ of 
$\cC$ such that $x\in\ri\,\tilde{F}(x)$. The inverse image 
$\tilde{G}(x):=\bE|_{\cM_\cA}^{-1}(\tilde{F}(x))$ is a face of the state space 
$\cM$ and we have  $\bE(\ri\,\tilde{G}(x))=\ri\,\tilde{F}(x)$. The face 
$\tilde{G}(x)$ is the state space of the C*-algebra $p\cA p$ for a unique 
orthogonal projection $p=p(x)\in\cA$ and $\ri\,\tilde{G}(x)$ consists of all 
invertible states in $p\cA p$. We define $\cP({\bf u})$ as the set of all 
orthogonal projections arising in this construction from points $x\in\cC$. 
Writing $p(F):=p(x)$ for $F=\tilde{F}(x)$, we have
\begin{equation}\label{def:face-projection}
\cP({\bf u})=\{p(F)\mid  F\text{ is a non-empty face of $\cC$}\}.
\end{equation}
We have shown in \cite{Weis-support} that the lattice of faces of $\cC$ is 
isomorphic to $\cP({\bf u})\cup\{0\}$.
\par
Knowing the support projections $\cP({\bf u})$ makes it easier to discuss the 
continuity of the inference $\Psi$. 
\begin{Exa}\label{exa:open-m3x3}
We show that the inference $\Psi$ in the algebra $\cA:={\rm Mat}(3,\bC)$
with respect to the observables from Example~\ref{exa:Knauf} is discontinuous 
for almost all ranking functions. We have computed in Section~3.3 of 
\cite{Weis-support}
\begin{equation}\label{eq:projections-staffelberg}
\cP({\bf u})=\{\rho(\alpha)\mid\alpha\in\,]0,2\pi[\,\}
\cup\{p,\id_3\}
\end{equation}
where $\rho(\alpha)$ is defined in Example~\ref{exa:Knauf}, 
$\rho(0)=\tfrac 1{2}(\id_2+\sigma_2)\oplus 0$ and $p:=\rho(0)+0_2\oplus 1$. 
The convex support $\cC$ is the unit disc by Example~\ref{exa:Knauf} but the 
state space $\cM=\cM_\cA$ is not a cone as in the case of the direct sum 
algebras in Example~\ref{exa:typical1}. In particular, the face 
$p\cM p=\bE|_\cM^{-1}(0,1)$ is not a segment but a three-dimensional Bloch 
ball. We write $x\in\rb(\cC)$ as $x(\alpha)=(\sin(\alpha),\cos(\alpha))$ 
for $\alpha\in[0,2\pi[\,$. Then $x(\alpha)=(0,1)$ corresponds to $\alpha=0$. 
For $\alpha>0$ the fiber $\bE|_\cM^{-1}(x(\alpha))$ is by 
(\ref{eq:projections-staffelberg}) equal to 
$\{\rho(\alpha)\}=\rho(\alpha)\cA\rho(\alpha)$, so 
$\Psi(x(\alpha))=\rho(\alpha)$ holds. Hence $\Psi$ is continuous at 
$\alpha(0)=(0,1)$ if and only if $\Psi(0,1)=\rho(0)$. For all other values 
of $\Psi(0,1)$ in the Bloch ball $\bE|_\cM^{-1}(0,1)$ the inference $\Psi$ 
is discontinuous at $(0,1)$. This, in the sense of Example~\ref{exa:typical1}, 
means that a discontinuity of $\Psi$ is typical. We mention that $\Psi$ is 
always continuous on $\cC\setminus\{(0,1)\}$ by 
Theorem~\ref{thm:openness-condition} applied to
Corollary~\ref{cor:ri_projection} and Lemma~\ref{lem:one-point-fibers}.
\end{Exa}
%
%
%
%
\subsection{Unions of Exponential Families}
\label{sec:unions-of-exponential-families}
\par
We introduce exponential families of arbitrary support and define an 
extension $\ext$ as a union of exponential families. The extension 
$\ext$ is the first description (\ref{eq:inference-ext}) of the set of 
ME-inference states $\Psi(\cC)$. We add a coordinate system
(\ref{eq:inv-temp}) of generalized inverse temperatures to 
(\ref{eq:inference-ext}). By definition the ME-inference $\Psi_\theta$ is 
point-wise an information projection of the prior to a linear constraint 
set. So we start with information projections from the outset. The 
analysis of the \I-projection to linear sets of states leads to a 
question about independence of the prior.
\begin{Def}[Information projections]
\label{def:information-projections}
Let $\rho\in\cM$ and $X\subset\cM$. We write 
$S(X,\rho):=\inf_{\sigma \in X}S(\sigma,\rho)$. If $S(X,\rho)<\infty$ and if 
a unique $\omega\in X$ exists such that $S(X,\rho)=S(\omega,\rho)$ then 
$\omega$ is the {\it \I-projection} of $\rho$ to $X$. We write 
$S(\rho,X):=\inf_{\sigma\in X}S(\rho,\sigma)$. If $S(\rho,X)<\infty$ and if a 
unique $\omega\in X$ exists such that $S(\rho,X)=S(\rho,\omega)$ then 
$\omega$ is the {\it \rI-projection} of $\rho$ to $X$. A sequence of states 
$(\sigma_i)\subset\cM$ {\it \rI-converges} to $\omega\in\cM$ if 
$\lim_{i\to\infty}S(\omega,\sigma_i)=0$. If every sequence 
$(\sigma_i)\subset X$ such that $\lim_{i\to\infty}S(\rho,\sigma_i)=S(\rho,X)$ 
\rI-converges, independent of the sequence, to a unique $\omega\in\cM$, not 
necessarily in $X$, then $\omega$ is the {\it generalized rI-projection} of 
$\rho$ to $X$.
\end{Def}
\par
Definition~\ref{def:information-projections} is an analogue from probability 
theory. We refer to \cite{CM03} about basic properties of these concepts 
which also apply to the present non-commutative setting. The following two 
definitions are based on observables ${\bf u}=(u_1,\ldots,u_k)$ and convex 
support $\cC=\bE(\cM)$ from Definition~\ref{def:entropic-inference}. 
\begin{Def}[Linear families]
\label{def:linear_families}
The {\it linear family} with {\it expected value} $x\in\cC$ is defined by
$\cL=\cL_x:=\{\rho\in\cM\mid\bE_{\bf u}(\rho)=x\}$.
\end{Def}
\par
By definition, for all $x\in\cC$ the ME-inference $\Psi_\theta(x)$ is the 
\I-projection of the prior $e^\theta/\tr(e^\theta)$ to $\cL_x$. The goal is
to get explicit expressions of these \I-projections.
\begin{Def}[Extension of an exponential family]
\label{def:exponential_families}
We assume $p\in\cA$ is a non-zero orthogonal projection. The 
{\it exponential family} with {\it support} $p$ is defined by 
\[
\cE
=
\cE_p
=
\cE_{p,\theta}
:=
\{\tfrac{pe^{p\nu p}}{\tr(pe^{p\nu p})}\mid 
\nu=\theta+\sum_{i=1}^k\lambda_iu_i, \lambda_i\in\bR, i=1,\ldots,k\}.
\]
A set of orthogonal projections $\cP({\bf u})$ is constructed in 
(\ref{def:face-projection}) from observables ${\bf u}$ and algebra $\cA$. 
We define the {\it extension}
\[
{\rm ext}(\cE_{\id}):=\bigcup_{p\in\cP({\bf u})}\cE_p.
\]
When $p\neq\id$ then we define ${\rm ext}(\cE_p)$ in the algebra $p\cA p$. 
Thereby we write $p{\bf u}p:=(pu_1 p,\ldots,pu_k p)$ and we construct a
set of projections $\cP(p{\bf u}p)$ with respect to the algebra $p\cA p$ 
and the convex support $\bE_{p{\bf u}p}(\cM_{p\cA p})\subset\bR^k$,
\[
{\rm ext}(\cE_p):=\bigcup_{q\in\cP(p{\bf u}p)}\cE_q.
\]
\end{Def}
\par
We remark that the extension ${\rm ext}(\cE_\id)$ is a super-set of
$\cE_\id$ because the equality $\cM_\cA=\bE|_{\cM_\cA}^{-1}(\cC)$ shows 
$\id\in\cP({\bf u})$. Similarly $\cE_p\subset{\rm ext}(\cE_p)$ holds for 
arbitrary non-zero projections $p\in\cA$. 
\par
Returning to the ME-inference $\Psi_\theta$ let us consider the exponential 
family $\cE:=\cE_\id$ of full support. A description of $\Psi_\theta(\cC)$ 
can be proved in two steps. First, $\bE|_{\ext}:\ext\to\cC$ is a bijection. 
This is proved in Lemma~6.9 in \cite{Weis-topology}, using the mean value 
parametrization (\ref{eq:mean-value-par}) and the stratification of the 
convex support into its faces, explained in the paragraph of 
(\ref{def:face-projection}). Therefore a map
\begin{equation}\label{def:geo-projection}
\Pi_\cE:\cM_\cA\to\ext,\qquad
\rho\mapsto \cL_{\bE(\rho)}\cap\ext
\end{equation}
is defined by identifying the one-element set $\cL_{\bE(\rho)}\cap\ext$ with 
its element. Secondly, Theorem~6.12 in \cite{Weis-topology} shows for all 
states $\rho\in\cM_\cA$, for $\cE=\cE_\id$ and for $\sigma\in\ext$, that the 
relative entropy satisfies the {\it Pythagorean identity}
\begin{equation}\label{eq:Pythagoras}
S(\rho,\sigma)=S(\rho,\Pi_\cE(\rho))+S(\Pi_\cE(\rho),\sigma).
\end{equation}
The non-trivial part of (\ref{def:geo-projection}) and (\ref{eq:Pythagoras}) is a 
possibly empty intersection $\cL\cap\cE$. Using $\sigma=e^\theta/\tr(e^\theta)$
in (\ref{eq:Pythagoras}), and the fact that $S(\rho,\Pi_\cE(\rho))\geq 0$ is zero 
only for $\rho=\Pi_\cE(\rho)$, we get for any $\rho\in\cM$
\begin{equation}\label{eq:inference-projection}
\Psi_\theta\circ\bE(\rho)=\Pi_\cE(\rho).
\end{equation}
Evaluating (\ref{eq:inference-projection}) for all $\rho\in\cM$, we arrive at the 
first description of the set of ME-inference states
\addtocounter{equation}{-1}
\begin{equation}\label{eq:inference-ext}
\Psi_\theta(\cC)=\ext.\tag{D1}
\end{equation}
\addtocounter{equation}{+1}%
\par
A second way to prove the Pythagorean identity (\ref{eq:Pythagoras}) is to use 
Csisz\'ar's method of a ``spherical geometry of'' the relative entropy, by 
generalizing Lemma 2.1 in \cite{Csiszar75} to a non-commutative algebra. A
proof of (\ref{eq:inference-ext}) using Lagrangian multipliers is included in
the PhD thesis \cite{Weis-Diss}, while the extended Pythagorean identity 
(\ref{eq:Pythagoras}) has first appeared in \cite{Weis-topology}.
\par
{\it Generalized inverse temperatures} have been introduced by Ingarden et al.\ 
\cite{Ingarden} as a physically motivated parametrization of a Gibbsian family 
$\cE=\cE_{\id,0}$. The adjective {\it generalized} refers to a number of $k>1$ 
observables. Generalized inverse temperatures are dual to expected values in 
a sense of information geometry, see Section~7.2 in the book 
\cite{Amari-Nagaoka} by Amari and Nagaoka and see Jen\v cov\'a \cite{Jencova}
for details. We have extended this coordinate system to $\ext$ by including a 
support projection in $\cP({\bf u})$, defined in 
Section~\ref{sec:progress-convex}, as a new parameter \cite{Weis-topology}. 
Now we generalize to arbitrary $\theta$. 
\begin{Thm}[Generalized inverse temperatures]
\label{thm:min-rel-ent}
Let $x=(\xi_1,\dots,\xi_k)\in\cC$. There exists a unique orthogonal projection 
$p\in\cP({\bf u})$ and there exist (generally non-unique) real numbers 
$\beta_1,\ldots,\beta_k$, such that 
$\nu(\beta):=\theta-\sum_{i=1}^k\beta_iu_i$ solves 
\addtocounter{equation}{-1}
\begin{equation}\label{eq:inv-temp}
\tfrac{\partial}{\partial\beta_j}\,
\log\tr(p e^{p\nu(\beta)p})=-\xi_j,
\qquad j=1,\ldots,k.\tag{D2}
\end{equation}
\addtocounter{equation}{+1}%
Each solution $(p,\beta_1,\ldots,\beta_k)\in\cP({\bf u})\times\bR^k$ of 
(\ref{eq:inv-temp}) defines a density matrix equal to the ME-inference
of $x$, that is $\Psi_\theta(x)=pe^{p\nu(\beta)p}/\tr(pe^{p\nu(\beta)p})$
holds. The relative entropy of $\Psi_\theta(x)$ from the prior 
$\sigma=e^\theta/\tr(e^\theta)$ is
\[
\min\{S(\rho,\sigma)\mid\rho\in\cL_x\}
=S(\Psi_\theta(x),\sigma)
=\log\tr e^\theta-\log\tr(pe^{p\nu(\beta)p})-\sum_{i=1}^k\beta_i\xi_i.
\]
\end{Thm}
{\em Proof:\/}
Corollary~6.11 in \cite{Weis-topology} shows that for every $x\in\cC$ there is
a tuple $t:=(p,\beta_1,\ldots,\beta_k)\in\cP({\bf u})\times\bR^k$ solving 
(\ref{eq:inv-temp}) with $p$ unique. Moreover, any state 
$pe^{p\nu p}/\tr(pe^{p\nu p})$ corresponding to a solution $t$ is equal to the 
unique state in $\omega\in\ext$ such that $\bE(\omega)=x$. Therefore 
(\ref{eq:inference-projection}) shows $\omega=\Psi_\theta(x)$. The 
computation of the relative entropy is straight forward. 
\hspace*{\fill}$\Box$\\
\par
In order to maximally extend the \I-projection to linear families, we 
rewrite the Pythagorean identity to allow for exponential families 
$\cE=\cE_p$ of arbitrary support $p$.  For subsets $X,Y\subset\cM$ let 
$S(X,Y):=\inf_{\rho\in X,\sigma\in Y}S(\rho,\sigma)$.
\begin{Thm}[Pythagorean identity]
\label{thm:projections}
Let $\cL\subset\cM_\cA$ be a linear family and $\cE\subset\cM_\cA$ be an 
exponential family such that $S(\cL,\cE)<\infty$. Then there is a unique state 
$\omega\in\cL\cap\ext$ and this state $\omega$ satisfies the
Pythagorean identity
\begin{equation}
\label{eq:pythagoras}
S(\rho,\sigma)=S(\rho,\omega)+S(\omega,\sigma),
\qquad
\rho\in\cL,\,
\sigma\in\ext.
\end{equation}
\end{Thm}
{\it Proof:}
Let $\cE=\cE_p$ for a non-zero orthogonal projection $p$ and let $\cL=\cL_x$ 
for an expected value $x\in\cC$. Let us find the intersection point of $\cL$ 
and $\ext$ in the *-algebra $p\cA p$. Since $S(\cL,\cE)<\infty$, the linear 
family $\cL$ intersects $\cM_{p\cA p}$ and by cyclic reordering under the 
trace $\cL_x\cap\cM_{p\cA p}$ is the same set for both observables 
${\bf u}=(u_1,\ldots,u_k)$ and  $p{\bf u}p=(pu_1p,\ldots,pu_kp)$. The linear 
map $\bE_{p{\bf u}p}:(p\cA p)_{\rm sa}\to\bR^k$ restricts, according to 
(\ref{def:geo-projection}) to a bijection between the extension $\ext$ and 
the convex support $\bE_{p{\bf u}p}(\cM_{p\cA p})\subset\bR^k$. Thus the 
linear family $\cL_x\cap\cM_{p\cA p}$ intersects the extension $\ext$ in a 
unique point, denoted by $\omega$. The Pythagorean identity 
(\ref{eq:Pythagoras}) proves for all $\rho\in\cL_x\cap\cM_{p\cA p}$ and for all 
$\sigma\in\ext$ the equality of  $S(\rho,\sigma)=S(\rho,\omega)+S(\omega,\sigma)$. 
This equality holds also for $\rho\in\cL\setminus\cM_{p\cA p}$ where for all 
$\sigma\in\ext$ we have $S(\rho,\sigma)=S(\rho,\omega)=\infty$.
\hspace*{\fill}$\square$\\
\par
The state $\omega$ in Theorem~\ref{thm:projections} is also characterized in 
terms of optimality. Taking in (\ref{eq:pythagoras}) the minimum 
over $\rho\in\cL$ we have 
\begin{equation}\label{eq:I-min}
S(\cL,\sigma)=S(\omega,\sigma),
\qquad \sigma\in\ext,
\end{equation}
because $\omega\in\cL$. Similarly, since $\omega\in\ext$, minimizing over
$\sigma\in\ext$ gives
\begin{equation}\label{eq:rI-min-rIe}
S(\rho,\ext)=S(\rho,\omega),
\qquad \rho\in\cL.
\end{equation} 
In particular, for all $\rho\in\cM_\cA$ such that $S(\rho,\cE)<\infty$, the
\rI-projection of $\rho$ to $\ext$ exists. More specifically if $\cE=\cE_p$ 
and if $\rho\in p\cA p$ then the \rI-projection of $\rho$ to $\ext$ is the
unique state in $\cL_{\bE(\rho)}\cap\ext$. This follows from 
(\ref{eq:pythagoras}) and (\ref{eq:rI-min-rIe}). The inequalities in 
(\ref{eq:ineqality-def-I}) and (\ref{eq:ineqality-def-rI}) are never strict.
They are to support the discussion in Section~\ref{sec:commutative}.
\begin{Cor}[\I-projection]
\label{cor:I-properties}
Let $S(\cL,\sigma)<\infty$ for a state $\sigma\in\cM_\cA$ and a linear family 
$\cL\subset\cM_\cA$. The \I-projection of $\sigma$ to $\cL$ exists and equals 
the unique state $\omega\in\cL\cap\ext$ for $\cE:=\cE_{p,\theta}$. Here 
$p:=s(\sigma)$ is the support projection of $\sigma$ and  
$\theta:=\log(\sigma)$ is defined in the algebra $p\cA p$.
Moreover, $\omega$ is characterized as follows.
\begin{enumerate}
\item
Any sequence $(\rho_i)\subset\cL$ with 
$S(\rho_i,\sigma)\stackrel{i\to\infty}{\rightarrow}S(\cL,\sigma)$
\I-converges to $\omega$. 
\item
The state $\omega$ is the unique state in $\cM_\cA$ solving the set of 
inequalities
\begin{equation}
\label{eq:ineqality-def-I}
S(\rho,\sigma)\geq S(\rho,\omega)
+S(\cL,\sigma),\qquad\rho\in\cL.
\end{equation}
\end{enumerate}
\end{Cor}
{\it Proof:}
This follows from Theorem~\ref{thm:projections} and (\ref{eq:I-min}).
\hspace*{\fill}$\square$\\
\par
See Shirokov \cite{Shirokov-charI}, Section~3, for related non-commutative
results in infinite dimensions. Before discussing independence of the prior, 
we study \I-projections.
\begin{Lem}[Range of the \I-projection]
\label{lem:I-projection-range}
If $\cL\subset\cM_\cA$ is a linear family then the set of \I-projections 
to $\cL$ of all in $\cA$ invertible states is the relative interior 
$\ri(\cL)$. 
\end{Lem}
{\it Proof:}
Let $x\in\cC$ and $\cL=\cL_x$. In the paragraph of (\ref{def:face-projection})
a face $\tilde{G}(x)$ of the state space $\cM_\cA$ is defined, such that the 
affine space $\bA:=\{a\in\sa\mid\bE(a)=x\}$ meets the relative interior 
$\ri\,\tilde{G}(x)$ and such that $\cL=\bA\cap\tilde{G}(x)$ holds. A standard
argument of convex geometry, see Corollary~6.5.1 in \cite{Rockafellar}, shows 
$\ri(\cL)=\bA\cap\ri\,\tilde{G}(x)=\cL\cap\ri\,\tilde{G}(x)$. Moreover, the 
face $\tilde{G}(x)$ has the form $\tilde{G}(x)=\cM_{p\cA p}$ for some 
$p\in\cP({\bf u})$ and so $\ri(\cL)=\cL\cap\ri(\cM_{p\cA p})$ holds. The 
relative interior $\ri(\cM_{p\cA p})$ consists by Proposition~2.9 in 
\cite{Weis-support} of all in $p\cA p$ invertible states. So we have for all 
$\rho\in\cL$ 
\begin{align*}
\rho\in\ri(\cL) &\iff \rho\in\ri\,\cM_{p\cA p}
\iff \rho=p\exp(p\theta p) \text{ for some } \theta\in\sa\\
&\iff \rho\in\ext \quad \text{for} \quad \cE=\cE_{\id,\theta}
\text{ and some } \theta\in\sa\\
&\iff \rho \text{ is the \I-projection of } e^\theta/\tr(e^\theta) 
\text{ to } \cL \text{ for some } \theta\in\sa.
\end{align*}
The equivalence in the second line follows from the definition of $\cP({\bf u})$
in (\ref{def:face-projection}) and from the 
Definition~\ref{def:exponential_families} of the extension $\ext$.
The equivalence in the last line follows from Corollary~\ref{cor:I-properties} 
about the \I-projection.
\hspace*{\fill}$\square$\\
\par
Example~\ref{exa:typical1} and Example~\ref{exa:open-m3x3} of state spaces 
$\cM_\cA$ suggest that the continuity of the ME-inference is independent of the 
prior.
\begin{Rem}\label{rem:stability}
We consider linear constraints on a convex body $K$ in the setting of
Definition~\ref{def:basic}. The assertion that the continuity of the maximizer
$H$ is independent of the objective functional $g$ translates through 
Theorem~\ref{thm:openness-condition} into the assertion that for all $w\in L$ 
the map $f|_K$ is open at a point of $F(w)$ if and only if $f|_K$ is open on 
$F(w)$. This form of independence is wrong in the examples mentioned above. 
We consider a weaker form of independence.
\begin{equation}
\label{eq:ideal-openness}
\forall w\in L:\hspace{5ex}
\parbox[t]{10cm}{%
If the restricted linear map $f|_K$ is open at a relative interior point
of $F(w)$ then $f|_K$ is open on $F(w)$.}
\end{equation}
If (\ref{eq:ideal-openness}) holds for state spaces $\cM_\cA$ then the 
continuity of the ME-inference $\Psi_\theta$ is independent of the prior 
$\theta$. To see this it suffices to consider 
Lemma~\ref{lem:I-projection-range}.
\par
We begin the discussion by noticing that 
(\ref{eq:ideal-openness}) is wrong for the class of affine sections of state 
spaces $\cM_\cA$. A three-dimensional section $K$ of $\cM_\cA$ for the algebra 
$\cA:={\rm Mat}(3,\bC)$ is described in Section~5.4 in \cite{Henrion} up to 
scaling by $1/3$. The convex body $K$ is the set of positive semi-definite 
matrices
\begin{equation}\label{eq:cayley}
\left(\begin{smallmatrix}1/3 & x & y\\x & 1/3 & z\\
y & z & 1/3\end{smallmatrix}\right)
\end{equation}
with real parameters $x,y,z$. This convex body $K$ looks like an inflated 
tetrahedron, it has six edges forming a regular tetrahedron with vertices 
$\tfrac{1}{3}((-1)^\alpha,(-1)^\beta,(-1)^\gamma)$ such that the sum of 
$\alpha,\beta,\gamma\in\{0,1\}$ is even. The rest of the relative boundary 
of $K$ is covered by one-point faces. The Zariski closure of the relative 
boundary of $K$ is known as the {\it Cayley cubic surface}. 
\par
The convex body $K$ defined in (\ref{eq:cayley}) violates (\ref{eq:ideal-openness}) 
because the boundary curvature of a planar affine section, perpendicular to an edge
of $K$, has a minimum along the given edge at the midpoint of that edge. We think
it is more than a coincidence that $K$ also violates the notion of stability 
mentioned in Section~\ref{sec:overview}: The union of all extremal points
of $K$ is not closed in the norm topology, so $K$ is not stable by a theorem
in \cite{Papadopoulou}. On the other hand, stability of state spaces $\cM_\cA$ 
is proved for example in Lemma~3 in \cite{Shirokov06}. Basic questions are: Does 
(\ref{eq:ideal-openness}) hold for state spaces $\cM_\cA$, does it hold for stable 
convex bodies? Are linear images of $\cM_\cA$ stable, does (\ref{eq:ideal-openness}) 
hold for them? 
\end{Rem}
%
%
%
%
\subsection{Geodesic Closures}
\label{sec:geodesic-closures}
\par
The $(\pm1)$-geodesic closure of an exponential family $\cE=\cE_\id$
was defined in Section~\ref{sec:overview} as a union of geodesics, each
with two limit points. We consider them as descriptions 
(\ref{eq:+1-closure}) respectively (\ref{eq:-1-closure}) of the set of 
ME-inference states $\Psi_\theta(\cC)$.
\par
The equality of the $(+1)$-geodesic closure of $\cE$ to
\addtocounter{equation}{-1}
\begin{equation}\label{eq:+1-closure}
\bigcup\{\cE_{p(F)}\mid F\text{ is a non-empty exposed face of $\cC$}\}
\tag{D3}
\end{equation}
\addtocounter{equation}{+1}%
is proved in Proposition~6.21 in \cite{Weis-topology}, where each projection
$p(F)$, defined in Section~\ref{sec:progress-convex}, gives rise to the face
$\cM_{p(F)\cA p(F)}=\bE|_{\cM_\cA}^{-1}(F)$ of $\cM_\cA$. The equality
(\ref{eq:+1-closure}) follows on one hand from the limit 
\[
\lim_{t\to\infty}e^{\theta+t u}/\tr(e^{\theta+t u})
=e^{p\theta p}/\tr(e^{p\theta p}),
\qquad \theta,u\in\sa,
\]
where the spectral projection $p:=p^+(u)$ corresponds to the largest spectral 
value of $u$. See also Proposition~10 in \cite{Weis-Knauf}. On the other hand
we have used in the proof a refinement of the lattice isomorphisms in 
Section~\ref{sec:progress-convex}, which characterizes exposed faces of $\cC$. 
According to (\ref{def:face-projection}) and the 
Definition~\ref{def:exponential_families} of $\ext$ we have
\[
\ext=\bigcup\{\cE_{p(F)}\mid F\text{ is a non-empty face of $\cC$}\}.
\]
Therefore the equality $\Psi_\theta(\cC)=\ext$ in (\ref{eq:inference-ext}) 
shows that (\ref{eq:+1-closure}) is a correct description of 
$\Psi_\theta(\cC)$ if and only if the convex support $\cC$ has no 
non-exposed faces.
\par
The {\it swallow family} in \cite{Weis-Knauf} is an example of a Gibbsian 
family where the $(+1)$-geodesic closure is strictly smaller than $\ext$. 
If the observables $u_1,\ldots,u_k$ are commutative then 
(\ref{eq:support-independence}) shows that the convex support 
$\cC$ is a polytope. A polytope has no non-exposed faces so the $(+1)$-geodesic 
closure of $\cE$ is a correct description of $\Psi_\theta(\cC)$ for commutative
observables.
\par
Equality between the $(-1)$-geodesic closure of $\cE$ and $\Psi_0(\cC)$ is 
shown in Theorem~25 in \cite{Weis-Knauf} for the {\it Staffelberg family} 
$\cE$ in Example~\ref{exa:Knauf}. This is a non-trivial example where the 
maximum-entropy inference $\Psi_0$ is discontinuous. Now we prove the 
statement in full generality.
\begin{Thm}\label{thm:-1-geodesic-closure}
For all observables $u_1,\ldots,u_k\in\sa$ and all $\theta\in\sa$ the 
$(-1)$-geodesic closure of the exponential family $\cE_{\id,\theta}$ equals 
the set of ME-inference states $\Psi_\theta(\cC)$.
\end{Thm}
{\it Proof:}
An unparametrized $(-1)$-geodesic in $\cE_{\id,\theta}$ is by definition 
the image of an open segment $s\subset\ri(\cC)$ under the ME-inference 
$\Psi_\theta$. Since the norm closure $\overline{s}$ is a closed segment, the 
restriction of the expected value functional $\bE$ to 
$F(\overline{s})=\bE|_\cM^{-1}(\overline{s})$ has the polytope $\overline{s}$ 
as its image. Then Corollary~\ref{cor:open_poly} shows that 
$\bE|_{F(\overline{s})}$ is open and Theorem~\ref{thm:openness-condition} 
shows that $\Psi_\theta|_{\overline{s}}$ is continuous. This shows
\addtocounter{equation}{-1}
\begin{equation}\label{eq:-1-closure}
\overline{\Psi_\theta(s)}
=\overline{\Psi_\theta|_{\overline{s}}(s)}
=\Psi_\theta|_{\overline{s}}(\overline{s})
\subset\Psi_\theta(\cC).
\tag{D4}
\end{equation}
\addtocounter{equation}{+1}%
Since the closures of open segments in $\ri(\cC)$ exhaust the convex support
$\cC$, the proof is complete.
\hspace*{\fill}$\square$\\
%
%
%
%
\subsection{Topological Closures}
\label{sec:topological-closures}
\par
The projection theorem, in its extended form (\ref{eq:projection-thm}), yields
three descriptions of the set of ME-inference states $\Psi_\theta(\cC)$. Unless 
stated otherwise we consider an exponential family $\cE$ of full support, that 
is $\cE=\cE_{\id,\theta}$. The norm closure $\overline{\cE}$ fits best into 
this collection of topological descriptions of $\Psi_\theta(\cC)$. We finish 
the section with applications of \I- and \rI-projections. 
\par
The {\it projection theorem}, Theorem~6.16 in \cite{Weis-topology}, shows for
$\cE=\cE_{\id,\theta}$ and all $\rho\in\cM_\cA$,
\begin{equation}
\label{eq:projection-thm}
\inf_{\sigma\in\cE}S(\rho,\sigma)=\min_{\sigma\in\ext}S(\rho,\sigma).
\end{equation}
Moreover, for each $\rho\in\cM_\cA$ the function $S(\rho,\cdot)$ has a unique 
local minimum on $\ext$ at $\Pi_\cE(\rho)$. Thereby $\Pi_\cE(\rho)$ was defined 
in (\ref{def:geo-projection}) as the unique state in $\ext$ having expected 
value $\bE(\rho)$.
\par
This theorem is proved using iterated limits of $(+1)$-geodesics in $\cE$ and 
in the extension $\ext$. We argue in Section~3.6 in \cite{Weis-topology} that 
a single limit is not sufficient because of differences between the 
$(+1)$-geodesic closure of $\cE$ and the extension $\ext$ which appear if 
the convex support $\cC$ has non-exposed faces, see the discussion in 
Section~\ref{sec:geodesic-closures}.
\par
The projection theorem (\ref{eq:projection-thm}) directly implies 
$\ext=\cl(\cE)$ for the rI-closure $\cl(\cE)$ defined in (\ref{def:cl-rI}). 
This equality, together with (\ref{eq:inference-ext}), shows
\addtocounter{equation}{-1}
\begin{equation}\label{eq:inference-closure}
\Psi_\theta(\cC)=\cl(\cE).\tag{D5}
\end{equation} 
\addtocounter{equation}{+1}%
Another description of $\Psi(\cC)$ follows because the \rI-closure of $\cE$ 
is the closure in the topology generated by the family of open sets
\addtocounter{equation}{-1}
\begin{equation}\label{eq:inference-topology}
\{\sigma\in\cM_\cA\mid S(\rho,\sigma)<\epsilon\},\tag{D6}
\qquad 
\rho\in\cM_\cA, \quad \epsilon>0.
\end{equation}
\addtocounter{equation}{+1}%
This family of subsets is the base of a topology on $\cM_\cA$, which we
call {\it rI-topology} in \cite{Weis-topology}. Theorem~5.18.2 in 
\cite{Weis-topology} proves that the \rI-closure of an arbitrary subset 
$X\subset\cM_\cA$ is the closure of $X$ in the \rI-topology. 
\par
Corollary~5.19 in \cite{Weis-topology} shows that the \rI-topology in a 
commutative algebra $\cA$ is the norm topology, restricted to $\cM_\cA$,
where the norm closure $\overline{\cE}$ equals $\Psi_\theta(\cC)$. More 
generally we have 
\addtocounter{equation}{-1}
\begin{equation}\label{eq:inference-norm-closure}
\Psi_\theta(\cC)=\overline{\cE}
\text{ if the observables } 
u_1,\ldots,u_k
\text{ commute.}\tag{D7}
\end{equation}
\addtocounter{equation}{+1}%
Clearly $\Psi_\theta(\cC)\subsetneq\overline{\cE}$ holds if $\Psi_\theta$ is 
not continuous, like in the Example~\ref{exa:Knauf}. If the observables 
$u_1,\ldots,u_k$ are commutative then (\ref{eq:support-independence}) shows 
that the convex support $\cC$ is a polytope. Corollary~\ref{cor:open_poly} 
shows that the expected value functional $\bE|_\cM$ is open and 
Theorem~\ref{thm:openness-condition} proves that $\Psi_\theta$ is continuous. 
\begin{Rem}
Csisz\'ar \cite{Csiszar67} has shown for infinite sigma-algebras that subsets of probability measures analogous to (\ref{eq:inference-topology}) do not define a 
topology. On the other hand the analogue of the \rI-convergence, see 
Definition~\ref{def:information-projections}, defines on every space of 
probability measures a topology which was investigated by Dudley 
\cite{Dudley98}. Harremo\"es \cite{Harremoes} has studied the corresponding 
topology of the \I-convergence. Our analysis of the \rI-topology 
(\ref{eq:inference-topology}) in \cite{Weis-topology} starts from convergence, 
too.
\end{Rem}
\par
Our final description of $\Psi_\theta(\cC)$ is a the generalized 
\rI-projection, see Definition~\ref{def:information-projections}, to the 
exponential family $\cE_{\id,\theta}$. First we study the maximal range. 
Let $\cE\subset\cM_\cA$ denote an exponential family of arbitrary support 
$p\in\cA$. We prove
\begin{equation}\label{eq:ext=cl}
\ext=\cl(\cE).
\end{equation}
While (\ref{eq:ext=cl}) holds in the algebra $p\cA p$ by
(\ref{eq:projection-thm}), we have to show that $\cl(\cE)$ does not 
increase if we enlarge the algebra from $p\cA p$ to $\cA$. But this is clear 
from $S(\rho,\cE)=\infty$ which holds for states $\rho\in\cM_\cA$ not in 
$p\cA p$. 
\par
We switch now from $\ext$ to $\cl(\cE)$ because the two sets are conceptually 
different. The extension $\ext$ is the set of \I-projections of a prior state 
to a class of parallel linear families. Corollary~\ref{cor:rI-properties} shows 
that the \rI-closure $\cl(\cE)$ is the range of the generalized \rI-projection 
to $\cE$. If $\cL$ is a linear family such that $S(\cL,\cE)<\infty$ then 
Theorem~\ref{thm:projections} shows the existence of a unique 
$\omega\in\cL\cap\cl(\cE)$. Taking in (\ref{eq:pythagoras}) the infimum over 
$\sigma\in\cE$ we get for all $\rho\in\cL$ by the definition (\ref{def:cl-rI}) 
of the \rI-closure
\begin{equation}\label{eq:rI-min-e}
S(\rho,\cE)=S(\rho,\omega).
\end{equation} 
\begin{Cor}[Generalized \rI-projection]
\label{cor:rI-properties}
Let $S(\rho,\cE)<\infty$ for a state $\rho\in\cM_\cA$ and an exponential 
family $\cE\subset\cM_\cA$. The \rI-projection of $\rho$ to $\cl(\cE)$
exists and equals the unique state $\omega\in\cL_{\bE(\rho)}\cap\cl(\cE)$. 
Moreover, $\omega$ is characterized as follows.
\begin{enumerate}
\item[1.]
Any sequence $(\sigma_i)\subset\cE$ with
$S(\rho,\sigma_i)\stackrel{i\to\infty}{\rightarrow} S(\rho,\cE)$ 
\rI-converges to $\omega$. 
\item[1'.]
Any sequence $(\sigma_i)\subset\cl(\cE)$  
$\lim_{i\to\infty}S(\rho,\sigma_i)\leq S(\rho,\cE)$ 
\rI-converges to $\omega$. 
\item[2.]
The state $\omega$ is the unique state in $\cM_\cA$ solving the set of 
inequalities
\begin{equation}
\label{eq:ineqality-def-rI}
S(\rho,\sigma)\geq S(\rho,\cE)+S(\omega,\sigma),
\qquad\sigma\in\cE.
\end{equation}
\end{enumerate}
\end{Cor}
{\it Proof:}
Using $\ext=\cl(\cE)$, proved in (\ref{eq:ext=cl}), this follows from 
Theorem~\ref{thm:projections}, (\ref{eq:rI-min-rIe}) and (\ref{eq:rI-min-e}).
\hspace*{\fill}$\square$\\
\par
Corollary~\ref{cor:rI-properties}.1 shows that the generalized \rI-projection 
of $\rho\in\cM_\cA$ to an exponential family $\cE\subset\cM_\cA$ exists, if 
$S(\rho,\cE)<\infty$, and then equals the \rI-projection to the \rI-closure 
$\cl(\cE)$. If $\cE$ has full support $\id$, then these maps are defined on 
$\cM_\cA$ and, using (\ref{eq:projection-thm}), 
\addtocounter{equation}{-1}
\begin{equation}\label{eq:generalized-rI}
\Pi_\cE \quad = \quad \text{generalized \rI-projection to }\cE
\tag{D8}
\end{equation}
\addtocounter{equation}{+1}%
holds for the map $\Pi_\cE$ defined in (\ref{def:geo-projection}), which by
(\ref{eq:inference-projection}) describes $\Psi_\theta(\cC)$. 
\begin{Rem}[Applications of information projections]~
\label{rem:projection-apps}
\begin{enumerate}
\item
The infimum $S(\cL,\rho)=\inf_{\sigma\in\cL}S(\sigma,\rho)$ is the minimal 
error probability in quantum hypothesis testing to decide for the linear
family $\cL$ while $\rho$ is the true state. In fact, a quantum version of 
Sanov's Theorem \cite{Bjelakovic-etal} shows that the minimal error 
probability is proportional to $e^{-nS(\cL,\rho)}$ for large $n$, if $n$ 
copies $\rho\otimes\cdots\otimes\rho$ of $\rho$ are accessible for
measurement. By Corollary~\ref{cor:I-properties} the infimum $S(\cL,\rho)$ 
is achieved at the \I-projection of $\rho$ to $\cL$ if $S(\cL,\rho)<\infty$.
\item
If the infimum $S(\rho_{AB},\cE)$ of the bipartite state $\rho_{AB}$ from
the exponential family 
\[
\cE=\{\sigma\otimes\tau\mid\sigma,\tau\text{ invertible states }\}
\]
of product states is achievable in $\cE$, then the \rI-projection of $\rho_{AB}$
to $\cE$ is the product $\rho_A\otimes\rho_B$ of the partial traces $\rho_A$, 
$\rho_B$ of $\rho_{AB}$ on the subsystems. So 
$S(\rho_{AB},\cE)=S(\rho_{AB},\rho_A\otimes\rho_B)$ equals the 
{\it mutual information} $S(\rho_A)+S(\rho_B)-S(\rho_{AB})$, defined in terms of 
the von Neumann entropy. The mutual information is a well-known measure of 
quantum correlations \cite{Groisman,Modi,Nielsen-Chuang}.
\end{enumerate}
\end{Rem}
%
%
%
%
\subsection{Literature From Probability Theory}
\label{sec:commutative}
\par
We discuss literature about information projections of probability measures 
in the context of the present article.
\par
We restrict the discussion to the finite sample space $\{1,\ldots,n\}$, 
$n\in\bN$ and sketch from this perspective some problems in infinite-dimensional 
spaces. Probability measures correspond to probability vectors in $\bR^n$ which 
will be identified with the states of the commutative C*-algebra 
$\cA_{\rm C}\subset{\rm Mat}(n,\bC)$ of diagonal matrices. As before, $\cA$ 
denotes a *-subalgebra of ${\rm Mat}(n,\bC)$.
\par
The essential advance from \v Cencov \cite{Cencov} to Csisz\'ar and 
Mat\'u\v s \cite{CM03} is the generalization of {\it convex support} to
{\it convex core} in the context of an exponential family 
of Borel probability measures on $\bR^d$. According to Section~I.C in 
\cite{CM03}, \v Cencov's results hold for finite support but are wrong for
infinite support where non-exposed faces of the convex core appear.
In the ``intersection'' between the two theories of Borel measures and 
finite-level quantum states lies the commutative algebra $\cA_{\rm C}$
where convex support and convex core coincide with our 
Definition~\ref{def:entropic-inference} of convex support which is a 
polytope. A polytope has no non-exposed faces.
\par
\v Cencov proved in 1972 an analogue of Theorem~\ref{thm:projections}  
to have a {\it maximum-likelihood estimation} defined with probability one. 
Given a probability measure $\rho\in\cA_{\rm C}$ and 
an exponential family $\cE\subset\cA_{\rm C}$ such that $S(\rho,\cE)<\infty$, 
Theorem 23.3 in \cite{Cencov} shows that the \rI-projection $\omega$ of $\rho$ 
to $\ext$ exists and is unique. \v Cencov also proves an inequality 
$S(\rho,\sigma)\geq S(\rho,\omega)+S(\omega,\sigma)$, $\sigma\in\ext$ of 
the form (\ref{eq:pythagoras}). A strict inequality is possible for Borel 
probability measures on $\bR$ by Remark~8 in \cite{CM03}.
\par
A second analogue of Theorem~\ref{thm:projections} was proved in 1975 by 
Csisz\'ar \cite{Csiszar75} aiming at a geometric understanding of the ME 
method. Some of his ideas extend to a non-com\-mu\-ta\-tive algebra $\cA$ 
and allow alternative proofs for the range of the mean value chart 
(\ref{eq:mean-value-par}) and the Pythagorean identity 
(\ref{eq:Pythagoras}) using only the derivative of the (matrix) logarithm 
and elementary calculus.
\par
Tops\o e \cite{Topsoe} has developed in 1979 a game theoretic foundation 
of the ME method. Some of his technical ideas extend to a 
non-com\-mu\-ta\-tive algebra and provide alternative proofs for 1.\ and 2.\ 
of Corollary~\ref{cor:I-properties}. Tops\oe's results, which apply to convex 
sets of measures rather than linear sets, and Csisz\'ar's in the previous
paragraph are based on the {\it parallelogram identity} of the relative 
entropy
\[
t S(\rho,\tau)+t' S(\sigma,\tau) = S(t\rho+t'\sigma,\tau)
+t S(\rho,t\rho+t'\sigma)+t' S(\sigma,t\rho+t'\sigma)
\]
which is easy to prove for $t\in[0,1], t'\equiv1-t$ and arbitrary states
$\rho,\sigma,\tau\in\cM_\cA$. See Shirokov \cite{Shirokov-charI}, Section~3, 
for related non-commutative results in infinite dimensions.
\par
Csisz\'ar and Mat\'u\v s have proved in 2003 a ``log-convex'' counterpart to 
Topsoe's ``convex'' results: The simplest special case of Theorem~1 in 
\cite{CM03} is included in Corollary~\ref{cor:I-properties} and 
Corollary~\ref{cor:rI-properties}, applied to $\cA_{\rm C}$. Moreover, an 
extensive analysis of exponential and linear families of Borel probability 
measures on $\bR^d$ is done in Theorem~3 and Theorem~4 in \cite{CM03}, which 
in the simple setting of $\cA_{\rm C}$ is included in 
Theorem~\ref{thm:projections} and in its substitute under $\cl(\cE)=\ext$.
%
%
%
%
\subsection{Divergence and the Norm Closure}
\label{sec:upper}
\par
We discuss the relative entropy from an exponential family $\cE$ because 
its continuity is intimately related to the continuity of the ME-inference 
$\Psi_\theta$ and because it characterizes the norm closure of the 
Gibbsian family in Example~\ref{exa:Knauf}. The relative entropy from 
$\cE$ generalizes the mutual information in a bipartite system in 
Remark~\ref{rem:projection-apps}.2 and was studied by Ay \cite{Ay02} and 
others as an abstract correlation measure. 
\begin{Def}[Divergence]\label{def:divergence}
We extend Definition~\ref{def:basic} by introducing a map $\Pi:K\to K$, 
$v\mapsto H\circ f(v)$ which we call {\it projection}, and a function
$\de:K\to\bR_0^+$, $v\mapsto h\circ f-g$ which we call {\it divergence}.
\end{Def}
\par
The image of the projection $\Pi$ is the set of maximizers $H(L)\subset K$, 
$\Pi$ preserves the fibers of $f:K\to L$ and 
$h\circ f=g\circ H\circ f=g\circ\Pi$ holds by definition. The divergence 
$\de$ quantifies how strong $g$ separates points on a fiber of $f$ from the 
maximizer in that fiber. In particular, a zero divergence $\de(v)=0$ of
$v\in K$ characterizes the set of maximizers $H(L)\subset K$, that is
$\de(v)=0\iff v=H\circ f(v)$. In the sequel we say a function is 
{\it continuous on} a subset of its domain if the function is continuous at 
each point of the given subset. For example, Dirichlet's function 
$\delta:\bR\to\{0,1\}$, $\delta(x)=0$ for $x\in\bQ$ and $\delta(x)=1$ for 
$x\in\bR\setminus\bQ$, is discontinuous on the rationals $\bQ$ but the 
restriction $\delta|_{\bQ}$ is continuous.
\begin{Lem}[Upper semi-continuity of the divergence]~
\label{lem:upper-semi-divergence}
\begin{enumerate}
\item
The divergence $\de:K\to\bR$ is upper semi-continuous on $K$. For all 
$v\in K$ the projection $\Pi:K\to K$ is continuous at $v$ if and only if
$\de$ is continuous at $v$.
\item
For all $w\in L$ the maximum $h$ is continuous at $w\in L$ if and only if 
$\de$ is continuous on the fiber $F(w)$. If $\de$ is continuous at 
$v\in K$ and $f|_K$ is open at $v$ then $h$ is continuous at $f(v)$.
\end{enumerate}
\end{Lem}
{\it Proof:}
In part one the upper semi-continuity of $\de=h\circ f-g$ follows from the 
continuity of $f$ and $g$ and the upper semi-continuity of $h$, proved in 
Lemma~\ref{lem:upper-semi-maximum}. The choices of $\tilde X:=K$ and 
$\tilde f:=f$ in that lemma show that the continuity of $h\circ f$ at 
$v\in K$ implies the continuity of $\Pi$ at $v$. Hence the continuity of
$\de=h\circ f-g$ at $v\in K$ implies the continuity of $\Pi$ at $v$. The 
converse follows from $\de=g\circ\Pi-g$.
\par
In part two the equality $\de=h\circ f-g$ shows that continuity of $h$ at
$w\in L$ implies the continuity of $\de$ on the fiber $F(w)$. In the other 
direction we first prove that the set-valued map $F$ is {\it closed} in the 
sense that for all $w_0\in L$ and $v_0\in K$ such that $v_0\not\in F(w_0)$, 
there are neighborhoods $N_1$ of $w_0$ and $N_2$ of $v_0$ such that 
$F(N_1)\cap N_2=\emptyset$. Using $w_1:=f(v_0)$, where $w_1\neq w_0$, and 
disjoint neighborhoods $N_1$ and $N_3$ of $w_0$ and $w_1$, respectively, it 
suffices to set $N_2:=F(N_3)=f|_K^{-1}(N_3)$. Since $F$ is closed and $K$ is 
compact, the Corollary to Theorem~7 in Section~VI.1 in \cite{Berge} proves 
for every $w\in L$ and every subset $N\subset K$ which is a neighborhood of 
all points in $F(w)$ that $f(N)$ is a neighborhood of $w$. Assuming the 
continuity of $h\circ f$ on $F(w)$, for every neighborhood $N\subset\bR$ of 
$h(w)$ the set $(h\circ f|_K)^{-1}(N)$ is a neighborhood of all points in 
$F(w)$. It follows that $h^{-1}(N)=f\circ(h\circ f|_K)^{-1}(N)$ is a 
neighborhood of $w$ and hence $h$ is continuous at $w$. The equality of 
set-valued maps $h^{-1}=f\circ(h\circ f|_K)^{-1}$ proves also the second 
statement of part two.
\hspace*{\fill}$\square$\\
\par
It is possible in Lemma~\ref{lem:upper-semi-divergence}.2 that $\de$ is 
continuous and discontinuous respectively on non-empty subsets of the 
same fiber of $f|_K$. Interestingly, such a disorder characterizes in 
Theorem~\ref{thm:d-continuity} the norm closure of a Gibbsian family. We
return to the setting of the ME-inference $\Psi_\theta$ in 
Definition~\ref{def:entropic-inference} and to exponential families 
$\cE=\cE_{\id,\theta}$ of full support, introduced in 
Definition~\ref{def:exponential_families}. We consider the function
\[
\de_\cE:\cM\to\bR,
\quad
\rho\mapsto S(\rho,\cE)
=\inf\{S(\rho,\sigma)\mid\sigma\in\cE\},
\]
which we had called {\it entropy distance} in \cite{Weis-Knauf}. The projection 
theorem (\ref{eq:projection-thm}) shows for all $\rho\in\cM$ that
\begin{equation}\label{eq:de-ext}
\de_\cE(\rho)
=S(\rho,\ext)
=S(\rho,\Pi_\cE(\rho))
\end{equation}
holds with $\Pi_\cE:\cM\to\ext$ introduced in (\ref{def:geo-projection}). 
We have shown in (\ref{eq:generalized-rI}) that $\Pi_\cE$ is the
generalized \rI-projection to $\cE$. 
\par
The Pythagorean identity (\ref{eq:Pythagoras}) and (\ref{eq:de-ext}) 
show $\de_\cE=\phi_\theta\circ\Pi_\cE-\phi_\theta$, while the equality 
$\psi_\theta\circ\bE|_\cM=\phi_\theta\circ\Pi_\cE$ follows from 
$\Psi_\theta\circ\bE=\Pi_\cE$ in (\ref{eq:inference-projection}). 
The equation $\de_\cE=\psi_\theta\circ\bE-\phi_\theta$ shows that $\de_\cE$ is 
a divergence in the sense of Definition~\ref{def:divergence}. The upper 
semi-continuity of the infimum $\de_\cE$ in Lemma~\ref{lem:upper-semi-divergence}.1 
follows also from the continuity of $\rho\mapsto S(\rho,\sigma)$ for the 
invertible states $\sigma\in\cE$ because a point-wise infimum preserves upper 
semi-continuity.
\par
We now specialize to the Staffelberg family $\cE$ in Example~\ref{exa:Knauf} and 
the algebra $\cA^{\rm Cone}$. The state space $\cM^{\rm Cone}$ of $\cA^{\rm Cone}$ 
is a three-dimensional cone. The observables are $u_1=\sigma_1\oplus 0$ and 
$u_2=\sigma_2\oplus 1$. We use the parametrization of $\cE$ by
\[
R: \bR^2\to\cM^{\rm Cone},\quad
(s,t)\mapsto\exp(s u_1+t u_2)/\tr[\exp(s u_1+t u_2)].
\] 
The convex support $\cC=\bE(\cM^{\rm Cone})$ is the closed unit disk in 
$\bR^2$ and the real analytic diffeomorphism 
$\bE\circ R:\bR^2\to\ri(\cC)\subset\bR^2$ in (\ref{eq:mean-value-par}) maps 
the plane onto the open unit disk. For $s\in\bR$ we consider $(+1)$-geodesics,
see Section~\ref{sec:overview}, and their expected values
\begin{equation}\label{def:e-geodesic_expected}
g_s:\bR\to\cE,\quad
t\mapsto R(s,t)
\qquad\text{ and }\qquad
h_s:\bR\to\ri(\cC),\quad
t\mapsto \bE\circ g_s(t),
\end{equation}
and we define 
$\gamma_s:=h_{-s}(\bR)\cup h_s(\bR)\cup\{(0,\pm1)\}\subset\cC$, $s\in\bR$.
We first prove some curvature estimates.
\begin{Lem}
\label{lem:projected-e-geodesic}
For each $s\in\bR$ we have $\lim_{t\to\pm\infty}h_s(t)=(0,\pm1)$. If $s>0$ then 
$\gamma_s\subset\cC$ is a simple closed curve, tangent to the unit circle at 
$(0,1)$, where $\gamma_s$ has curvature two, independent of $s$.
\end{Lem}
{\it Proof:} 
The limit $\lim_{t\to\pm\infty}h_s(t)=(0,\pm 1)$, $s\in\bR$, is computed in 
Lemma~23 in \cite{Weis-Knauf}. Since $\bE\circ R:\bR^2\to\ri(\cC)$ is a 
diffeomorphism, all parameters $s>0$ define pairs of compact curves 
$h_s(\bR)\cup\{(0,\pm1)\}$ and $h_{-s}(\bR)\cup\{(0,\pm1)\}$ intersecting only 
in $(0,\pm1)$ so $\gamma_s$ is a simple closed curve. We 
write $h_s(t)=(x(t),y(t))$, $b:=\sqrt{s^2+t^2}$ and $\eta:=e^b+e^{-b}+e^t$. 
By (25) in \cite{Weis-Knauf} we have
\[
x(t)=\tfrac{1}{\eta}(e^b-e^{-b})\tfrac{s}{b}
\qquad\text{and}\qquad
y(t)=\tfrac{1}{\eta}((e^b-e^{-b})\tfrac{t}{b}+e^t).
\]
Taylor expansion at $t=\infty$ shows
\[
\lim_{t\to\infty}t\tfrac{\partial y}{\partial x}=-s,\quad
\lim_{t\to\infty}t^2\tfrac{\partial x}{\partial t}=-\tfrac{1}{2}s,\quad
\lim_{t\to\infty}t^3\tfrac{\partial^2 x}{\partial t^2}=s\quad\text{and}\quad
\lim_{t\to\infty}t^4\tfrac{\partial^2 y}{\partial t^2}=-\tfrac{3}{2}s^2.
\]
Then $\lim_{t\to\infty}\tfrac{\partial^2y}{\partial x^2}=-2$ and
$\lim_{t\to\infty}\tfrac{\partial y}{\partial x}=0$ show that the 
curvature is two.
\hspace*{\fill}$\square$\\
\begin{Lem}
\label{lem:projected-conic-section}
Let $C\subset\bR^{d+2}$, $d\geq 1$, be a cone (quadric), symmetric
under the orthogonal group $O(d+1)$, and with half-angle 
$\varphi\in(0,\tfrac{\pi}{2})$. Introducing orthogonal coordinates 
$(x_1,\ldots,x_d,y,z)$ we assume $C$ is given by the solutions of
\[
x_1^2 + \cdots + x_d^2 +
(y\cos(\varphi) + z\sin(\varphi))^2 = 
\tan(\varphi)^2(- y\sin(\varphi) + z\cos(\varphi))^2.
\]
Then $C$ has apex $0\in\bR^{d+2}$, the $z$-axis lies in $C$ and $C$ 
is $O(d)$-symmetric in the coordinates $x\equiv(x_1,\ldots,x_d)$. Let 
$H$ be the hyperplane intersecting the $z$-axis orthogonally in $z<0$. 
Then the conic section $C\cap H$ has a local parametrization 
$x\mapsto (x,y(x),z)$ and $y(x)$ has Hessian $-\cot(\varphi)/z\cdot\id_d$ 
at the critical point $x=0$.
\end{Lem}
{\it Proof:} 
The above equation simplifies to
\[
(1-\tan(\varphi)^2)y^2 + 2z\tan(\varphi)y + x_1^2 + \cdots + x_d^2 = 0.
\] 
The discriminant is positive for small $x_i$, $i=1,\ldots,d$, so $y$ is 
an analytic function of $x_1^2 + \cdots + x_d^2$ with leading term 
$-\tfrac{1}{2}(x_1^2 + \cdots + x_d^2)\cot(\varphi)/z$.
\hspace*{\fill}$\square$\\
\par
Since the Staffelberg family $\cE$ is a Gibbsian family, the set of 
maximum-entropy inference states $\Psi_0(\cC)$ is the \rI-closure 
$\cl(\cE)$, see (\ref{eq:inference-closure}). Let us recall $\cl(\cE)$ 
and the norm closure $\overline{\cE}$, computed in Theorem~18 in 
\cite{Weis-Knauf}, in terms of components from Example~\ref{exa:Knauf}. 
The \rI-closure $\cl(\cE)$ is the union of $\cE$ 
with $\{c\}$ and with the directrix of the cone $\cM^{\rm Cone}$, but 
without $\rho(0)$. The norm closure exceeds $\cl(\cE)$ exactly in the 
set $[\rho(0),c[\,=\{(1-\lambda)\rho(0)+\lambda c\mid\lambda\in[0,1[\,\}$. 
\begin{Thm}
\label{thm:d-continuity}
The entropy distance $\de_\cE:\cM^{\rm Cone}\to\bR$ from the Staffelberg family
$\cE$ is discontinuous at each point of $\overline{\cE}\setminus\cl(\cE)$ and
continuous on the complement of $\overline{\cE}\setminus\cl(\cE)$ in 
$\cM^{\rm Cone}$.
\end{Thm}
{\it Proof:} 
The entropy distance $\de_\cE$ is discontinuous at each point in 
$\overline{\cE}\setminus\cl(\cE)$ because $\de_\cE$ vanishes precisely on 
$\cl(\cE)$ by the definition of the \rI-closure in (\ref{def:cl-rI}). 
Continuity of the maximum-entropy inference $\Psi_0:\cC\to\cM^{\rm Cone}$ 
is shown in Example~\ref{exa:typical1} on $\cC\setminus\{x\}$ for $x:=(0,1)$. 
Lemma~\ref{lem:upper-semi-maximum} shows that the maximum $\psi_0$ is 
continuous on $\cC\setminus\{x\}$ and Lemma~\ref{lem:upper-semi-divergence}.2 
shows that the entropy distance $\de_\cE$ is continuous on 
$\cM^{\rm Cone}\setminus\cL_x$. It remains to study the continuity of 
$\de_\cE$ at all points of the linear family $\cL_x=[\rho(0),0_2\oplus 1]$ 
outside of $\overline{\cE}\setminus\cl(\cE)$. The \rI- and the norm closure 
of $\cE$ are recalled above. The \rI-closure contains $c\in\cL_x$. The 
non-negative function $\de_\cE$ is upper semi-continuous by 
Lemma~\ref{lem:upper-semi-divergence}.1 so $\de_\cE$ is continuous on $\cl(\cE)$
where $\de_\cE\equiv0$. It remains to prove that $\de_\cE$ is continuous at
each $\rho\in\,]c,0_2\oplus 1]$. 
\par
Using $(+1)$-geodesics we first show that $\Psi_0|_M$ is continuous on a 
certain subset $M\subset\cC$ defined as a union of expected values of 
$(+1)$-geodesics
\[
M:=\{\pm x\}\cup\bigcup_{s\in[-1,1]}h_s.
\]
Since $\Psi_0$ is continuous on $\cC\setminus\{x\}$ it suffices to prove
continuity of $\Psi_0|_M$ at $x$. Asymptotics of curves 
$h_s(t)=\bE\circ g_s(t)=\bE\circ R(s,t)$, defined in 
(\ref{def:e-geodesic_expected}) and studied in Lemma~24 in \cite{Weis-Knauf}, 
show for $|s|\leq 1$ that $t\cdot\|R(s,t)-c\|$ is bounded uniformly in $s$ 
for large $t$. Let $(x_i)_{i\in\bN}\subset M\setminus\{\pm x\}$ converge to 
$x$. The diffeomorphism $\bE\circ R:\bR^2\to\ri(\cC)$ in 
(\ref{eq:mean-value-par}) shows that there are real numbers $s_i,t_i$ such 
that $|s_i|\leq 1$ and such that $x_i=\bE(R(s_i,t_i))$ holds for $i\in\bN$.
We have $R(s_i,t_i)=\Psi_0(x_i)$, $i\in\bN$, because $\cE$ consists of 
maximum-entropy states by (\ref{eq:mean-value-par}). Since the limit $x$ of 
$(x_i)$ does not lie in $\ri(\cC)$, the continuity of $\bE\circ R$ implies 
$t_i\stackrel{i\to\infty}{\rightarrow}\infty$. So 
$\Psi_0(x_i)=R(s_i,t_i)\stackrel{i\to\infty}{\rightarrow}c$ follows and 
since $\Psi_0(x)=c$ holds by Example~\ref{exa:Knauf} this shows the continuity 
of $\Psi_0|_M$.
\par
Using simple curvature estimates we deduce from the continuity of $\Psi_0|_M$
the continuity of $\Psi_0|_e$ for certain filled ellipses $e$ in the unit disk
$\cC$. Using the direction $u_3=0_2\oplus 1-\rho(0)$ of the generatrix $\cL_x$
we consider the neighborhood of $\rho$
\[
N:=\{\tau\in\cM^{\rm Cone}\mid\langle \tau,u_3\rangle
\geq\langle \tfrac{1}{2}(c+\rho),u_3\rangle\}.
\]
The set of expected values $e:=\bE(N)$ is bounded near $x\in e$ by an ellipse.
To compare numerical curvature values we replace the observables $u_1,u_2$ in 
Example~\ref{exa:Knauf} and $u_3$ by an ONB of tangent vectors to the affine
hull of $\cM^{\rm Cone}$. We set $v_1:=\tfrac{1}{\sqrt{2}}u_1$, 
$v_2:=\tfrac{1}{2\sqrt{6}}(3u_2-\id_3)$ and ${\bf v}:=(v_1,v_2)$. Using
\[
a:\bR^2\to\bR^2,\quad
(x_1,x_2)\mapsto(\sqrt{2}x_1,\tfrac{1}{3}+\sqrt{\tfrac{8}{3}}x_2)
\]
we have for all self-adjoint matrices $b\in\cA^{\rm Cone}$ of trace one 
$\bE_{\bf u}(b)=a\circ\bE_{\bf v}(b)$. So the convex support $\cC_{\bf u}$
is translated and stretched relative to $\cC_{\bf v}$ which is isometric to 
the orthogonal projection of $\cM^{\rm Cone}$ along $u_3$. 
Lemma~\ref{lem:projected-conic-section} shows that the curvature of the 
relative boundary arc of $e$ through $x$ can be estimated by
\[
\cot(\tfrac{\pi}{6})/\|0_2\oplus 1-\tfrac{1}{2}(c+\rho)\|
\cdot\sqrt{\tfrac{8}{3}}/(\sqrt{2})^2
>
\sqrt{2}/\|0_2\oplus 1-c\|=2.
\]
The relative boundary curve of $M$ has curvature two at $x$ by 
Lemma~\ref{lem:projected-e-geodesic}. Thus there exists a neighborhood $N'$ 
of $x$ such that $e\cap N'\subset M\cap N'$. Since $\Psi_0|_M$ is continuous 
and $\Psi_0$ is continuous on $\cC\setminus\{x\}$ this shows that 
$\Psi_0|_e$ is continuous.
\par
Finally, we put all details together. The entropy distance $\de_\cE$ is upper 
semi-continuous by Lemma~\ref{lem:upper-semi-divergence}.1 so it suffices to 
prove that $\de_\cE$ is lower semi-continuous at $\rho$. For all states $\tau$ 
in the neighborhood $N$ of $\rho$ we have
\[
\de_\cE(\tau)
=
S(\tau,\ext)
\leq 
S(\tau,\Psi_0(e))
\leq 
S(\tau,\Psi_0\circ\bE(\tau))
=
\de_\cE(\tau)
\]
by (\ref{eq:de-ext}), (\ref{eq:inference-ext}), because $\bE(\tau)\in e$ 
and by (\ref{eq:inference-projection}) and (\ref{eq:de-ext}). Therefore 
$\de_\cE(\tau)=S(\tau,\Psi_0(e))$ holds for all $\tau\in N$ and it 
suffices to show that $\cM^{\rm Cone}\to\bR,\rho\mapsto S(\rho,\Psi_0(e))$ 
is lower semi-continuous. Since $e$ is compact and $\Psi_0|_e$ is continuous
it follows that $\Psi_0(e)$ is compact. The lower semi-continuity of the 
relative entropy \cite{Wehrl} and Remark~\ref{rem:compact-infimum-lower} 
complete the proof.
\hspace*{\fill}$\square$\\
\par
The characterization of $\overline{\cE}$ in Theorem~\ref{thm:d-continuity} 
in terms of continuity of $\de_\cE$ extends to the state space of the algebra 
$\cA:={\rm Mat}(2,\bC)\oplus\bC$ where Lemma~\ref{lem:projected-conic-section} 
and the symmetry of the cone $\cM_\cA$ lead to the same curvature estimate
as in $\cM^{\rm Cone}$. It is unclear to the author how to extend this 
characterization to ${\rm Mat}(3,\bC)$.
%
%
%
\vspace{3mm}
\begin{Ack}
I have the pleasure to thank my colleagues, 
Nihat Ay, Wolfgang L\"ohr and Arleta Szko\l a for comments on a draft of 
this article, Qi Ding for discussions about curvature, Andreas Knauf for 
the proof of Lemma~\ref{lem:projected-conic-section}, and Arleta Szko\l a 
for discussions about hypothesis testing and maximum entropy. I thank 
Prof.\ Shavkat Ayupov and Prof.\ Maksim E.\ Shirokov for the help with
real von Neumann algebras and information convergence, respectively. This 
work is financed by the DFG project 
{\it Quantum Statistics: Decision problems and entropic 
functionals on state spaces}.
\end{Ack}
%
%
%
%
\phantomsection
\addcontentsline{toc}{section}{References}
\bibliographystyle{nnapalike}

\begin{thebibliography}{99999}
%
\bibitem{Alfsen-Shultz} Alfsen, E.\,M.\ and Shultz, F.\,W.:
{\it State Spaces of Operator Algebras:\ Basic Theory, Orientations, 
and C*-Products}, Springer (2001)
%
\bibitem{Ali12} Ali, S.\,A.; Cafaro, C.; Giffin, A.; Lupo, C.\ 
and Mancini, S.:
{\it On a Differential Geometric Viewpoint of Jaynes' MaxEnt Method 
and its Quantum Extension}, AIP Conf.\ Proc.\ 1443, 120--128 (2012)
%
\bibitem{Amari} Amari, S.:
{\it Differential-Geometrical Methods in Statistics},
Lecture Notes in Statistics 28, Springer (1985)
%
\bibitem{Amari-Nagaoka} Amari, S.\ and Nagaoka, H.: 
{\it Methods of Information Geometry},
AMS Translations of Mathematical Monographs 191 (2000)
%
\bibitem{Ay02} Ay, N.: 
{\it An Information-Geometric Approach to a Theory of Pragmatic Structuring}, 
Annals of Probability 30, 416--436 (2002)
%
\bibitem{Ayupov97} Ayupov, Sh.; Rakhimov, A.\ and Usmanov, Sh.:
{\it Jordan, Real and Lie Structures in Operator Algebras},
Springer (1997)
%
\bibitem{Barndorff} Barndorff-Nielsen, O.: 
{\it Information and Exponential Families in Statistical Theory},
John Wiley \& Sons, New York (1978)
%
\bibitem{Berge} Berge, C.: 
{\it Topological Spaces},
Dover Publications, Inc., New York (1997)
%
\bibitem{Bjelakovic-etal} Bjelakovi\'c, I.; Deuschel, J.-D.; Kr\"uger, T.;
Seiler, R.; Siegmund-Schultze, R.\ and Szko\l a, A.:
{\it A Quantum Version of Sanov's Theorem},
Commun.\ Math.\ Phys.\ 260, 659--671 (2005)
%
\bibitem{Boltzmann77} Boltzmann, L.:
{\it \"Uber die Beziehung zwischen dem zweiten Hauptsatze der mechanischen
W\"armetheorie und der Wahrscheinlichkeitsrechnung respektive den S\"atzen
\"uber das W\"armegleichgewicht},
Wien.\ Ber.\ 76, 373--435 (1877)
%
\bibitem{Bratteli} Bratteli, O.\ and Robinson, D.\,W.\,W.:
{\it Operator Algebras and Quantum Statistical Mechanics 2: 
Equilibrium States. Models in Quantum Statistical Mechanics},
Springer (1997)
%
\bibitem{Buzek99} Bu\v zek, V.; Drobn\'y, G.; Derka, R.; Adam, G.\ 
and Wiedemann, H.:
{\it Quantum State Reconstruction from Incomplete Data},
Chaos, Solitons \& Fractals 10, 981--1074 (1999)
%
\bibitem{Caticha12} Caticha, A.:
{\it Entropic Inference and the Foundations of Physics}, 
Brazilian Chapter of the International Society for Bayesian 
Analysis-ISBrA, Sao Paulo, Brazil (2012)
%
\bibitem{Caticha-Giffin06} Caticha, A.\ and Giffin, A.:
{\it Updating Probabilities}, 
AIP Conf.\ Proc.\ 872, 31--42 (2006)
%
\bibitem{Cencov} \v{C}encov, N.\,N.:
{\it Statistical Decision Rules and Optimal Inference}, 
AMS Translations of Mathematical Monographs 53 (1982);
Original publication: Nauka (1972)
%
\bibitem{Chen12} Chen, J.; Ji, Z.; Ruskai, M.\,B.\ and Zeng, B.:
{\it Comment on Some Results of Erdahl and the Convex Structure of 
Reduced Density Matrices},
Journal of Mathematical Physics 53, 072203 (2012)
%
\bibitem{Csiszar67} Csisz\'ar, I.: 
{\it On Topological Properties of f-Divergences},
Studia Sci.\ Math.\ Hungar.\ 2, 329--339 (1967)
%
\bibitem{Csiszar75} Csisz\'ar, I.:
{\it $I$-Divergence Geometry of Probability Distributions and Minimization 
Problems}, Ann.\ Prob.\ 3, 146--158 (1975)
%
\bibitem{Csiszar91} Csisz\'ar, I.:
{\it Why Least Squares and Maximum Entropy? An Axiomatic Approach to Inference
for Linear Inverse Problems},
Ann.\ Statist.\ 19, 2032--2066 (1991)
%
\bibitem{CM03} Csisz\'ar, I.\ and Mat\'u\v s, F.:
{\it Information Projections Revisited}, 
IEEE Trans.\ Inf.\ Theory 49, 1474--1490 (2003)
%
\bibitem{Davies-Lewis70} Davies, E.\,B.\ and Lewis, J.\,T.:
{\it An Operational Approach to Quantum Probability},
Commun.\ Math.\ Phys.\ 17, 239--260 (1970) 
%
\bibitem{Dudley98} Dudley, R.\,M.:
{\it Consistency of M-Estimators and One-Sided Bracketing}, 
Progr.\ Probab.\ 43, 33--58, Birkh\"auser (1998)
%
\bibitem{Greenberger09} Greenberger, D.; Hentschel, K.\ and Weinert, F.:
{\it Compendium of Quantum Physics: Concepts, Experiments, History and 
Philosophy}, Springer (2009). 
%
\bibitem{Groisman} Groisman, B.; Popescu, S.\ and Winter, A.:  
{\it Quantum, Classical, and Total Amount of Correlations in a 
Quantum State}, Phys.\ Rev.\ A 72, 032317 (2005)
%
\bibitem{Harremoes} Harremo\"es, P.:
{\it Information Topologies with Applications}, 
Entropy,\! Search,\! Complexity, 
Bolyai Soc.\ Math.\ Stud.\ 16, 113--150 (2007)
%
\bibitem{Hasegawa} Hasegawa, H.:
{\it Exponential and Mixture Families in Quantum Statistics:\! 
Dual Structure and Unbiased Parameter Estimation},
Rep.\ Math.\ Phys.\ 39, 49--68 (1997)
%
\bibitem{Henrion} Henrion, D.: 
{\it Semidefinite Representation of Convex Hulls of Rational Varieties},
Acta Applicandae Mathematicae 115, 319--327 (2011)
%
\bibitem{Holevo73} Holevo, A.\,S.:
{\it Statistical Problems in Quantum Physics},
Lecture Notes in Mathematics 330, 104--119,
Springer (1973)
%
\bibitem{Ingarden} Ingarden, R.\,S.; Kossakowski, A.\ and Ohya, M.:
{\it Information Dynamics and Open Systems}, Kluwer Academic Publishers 
Group (1997)
%
\bibitem{Jaynes} Jaynes, E.\,T.:
{\it Information Theory and Statistical Mechanics}, 
Phys.\ Rev.\ 106, 620--630 (1957)
%
\bibitem{Jencova} Jen\v cov\'a, A.: 
{\it Geometry of Quantum States:\ Dual Connections and Divergence Functions}, 
Rep.\ Math.\ Phys.\ 47, 121--138 (2001)
%
\bibitem{Karbelkar86} Karbelkar, S.\,N.:
{\it On the Axiomatic Approach to the Maximum Entropy Principle of Inference},
Pram\=ana J.\ Phys.\ 26, 301--310 (1986)
%
\bibitem{Leung-Smith09} Leung, D.\ and Smith, G.:
{\it Continuity of Quantum Channel Capacities},
Commun.\ Math.\ Phys.\ 292, 201--215 (2009)
%
\bibitem{Li03} Li, B.:
{\it Real Operator Algebras},
World Scientific (2003)
%
\bibitem{Lima} Lima, \AA.:
{\it On Continuous Convex Functions and Split Faces},
Proc.\ London Math.\ Soc.\ 25, 27--40 (1972)
%
\bibitem{Modi} Modi, K.; Paterek, T.; Son, W.; Vedral, V.\ and 
Williamson, M.: 
{\it Unified View of Quantum and Classical Correlations},
Phys.\ Rev.\ Lett.\ 104, 080501 (2010)
%
\bibitem{Nagaoka} Nagaoka, H.:
{\it Differential Geometrical Aspects of Quantum State Estimation
and Relative Entropy}, Quantum Communication,\! Computing\! and\! 
Measurement, eds.\ Hirota et al., Plenum Press (1994)
%
\bibitem{Netzer-Plaumann-Schweighofer10} Netzer, T.; Plaumann, D.\
and Schweighofer, M.:
{\it Exposed Faces of Semidefinitely Representable Sets},
Siam J.\ Optim.\ 20, 1944--1955 (2010)
%
\bibitem{vonNeumann27} von Neumann, J.:
{\it Thermodynamik quantenmechanischer Gesamtheiten},
G\"ott.\ Nach.\ 273--291 (1927)
%
\bibitem{Nielsen-Chuang} Nielsen, M.\,A.\ and Chuang, I.\,L.: 
{\it Quantum Computation and Quantum Information},
Cambridge University Press (2000)
%
\bibitem{Papadopoulou} Papadopoulou, S.: 
{\it On the Geometry of Stable Compact Convex Sets},
Math.\ Ann.\ 229, 193--200 (1977)
%
\bibitem{Petz94} Petz, D.:
{\it Geometry of Canonical Correlation on the State Space of a Quantum 
System}, J.\ Math.\ Phys.\ 35, 780--795 (1994)
%
\bibitem{Petz08} Petz, D.: 
{\it Quantum Information Theory and Quantum Statistics}, 
Theoretical and Mathematical Physics, Springer (2008)
%
\bibitem{Protasov} Protasov, V.\,Yu.\ and Shirokov, M.\,E.:
{\it Generalized Compactness in Linear Spaces and its Applications},
Sbornik:\ Mathematics 200, 697--722 (2009)
%
\bibitem{Rockafellar} Rockafellar, R.\,T.:
{\it Convex Analysis},
Princeton University Press (1972) 
%
\bibitem{Ruskai88} Ruskai, M.\,B.: 
{\it Extermal Properties of Relative Entropy in Quantum Statistical 
Mechanics}, Rep.\ Math.\ Phys.\ 26, 143--150 (1988)
%
\bibitem{Schroedinger35} Schr\"odinger, E.:
{\it Die gegenw\"artige Situation in der Quantenmechanik},
Die Naturwissenschaften 23, 807--849 (1935)
%
\bibitem{Schweighofer} Schweighofer, M.; Sturmfels, B.\ and Thomas, R.:
{\it Convex Algebraic Geometry}, Banff Workshop Report (2010)
%
\bibitem{Shannon48} Shannon, C.\,E.: 
{\it A Mathematical Theory of Communication},
Bell System Technical Journal 27, 379--423 and 623--656 (1948)
%
\bibitem{Shirokov-charI} Shirokov, M.\,E.:
{\it Entropy Characteristics of Subsets of States. I},
Izvestiya: Mathematics 70, 1265--1292 (2006)
%
\bibitem{Shirokov06} Shirokov, M.\,E.:
{\it The Holevo Capacity of Infinite Dimensional Channels and the
Additivity Problem},
Commun.\ Math.\ Phys.\ 262, 137--159 (2006)
%
\bibitem{Shore-Johnson80} Shore, J.\,E.\ and Johnson, R.\,W.:
{\it Axiomatic Derivation of the Principle of Maximum Entropy and 
the Principle of Minimum Cross-Entropy},
IEEE Trans.\ Inf.\ Theory 26, 26--37 (1980); 
Correction {\it ibid.} 29, 942--943 (1983)
%
\bibitem{Skilling88} Skilling, J.:
{\it The Axioms of Maximum Entropy}, Maximum Entropy and Bayesian Methods,
eds.\ Erickson, G.\,J.\ and Smith, C.\,R., 173--187, 
Kluwer Academic Publishers (1988)
%
\bibitem{Topsoe} Tops\o e, F.: 
{\it Information Theoretical Optimization Techniques},
Kybernetika 15, 8--27 (1979)
%
\bibitem{Uffink95} Uffink, J.:
{\it Can the Maximum Entropy Principle be Explained as a Consistency 
Requirement?},
Stud.\ Hist.\ Phil.\ Sci.\ B 26, 223--261 (1995)
%
\bibitem{Uffink06} Uffink, J.:
{\it Compendium of the Foundations of Classical Statistical Physics},
Handbook for Philosophy of Physics, eds.\ Butterfield, J.\ and Earman, J.\ (2006) 
%
\bibitem{Uhlmann09} Uhlmann, A.:
{\it Roofs and Convexity}, Entropy 12, 1799--1832 (2010)
%
\bibitem{Umegaki} Umegaki, H.: 
{\it Conditional Expectation in an Operator Algebra, IV}, 
K\=odai Math.\ Sem.\ Rep.\ 14, 59--85 (1962)
%
\bibitem{Voigt} Voigt, I.\ and Weis, S.: 
{\it Polyhedral Voronoi Cells},
Contrib.\ Algebra and Geometry 51, 587--598 (2010) 
%
\bibitem{Wehrl} Wehrl, A.: 
{\it General Properties of Entropy},
Reviews of Modern Physics 50, 221--260 (1978)
%
\bibitem{Weis-Diss} Weis, S.:
{\it Exponential Families with Incompatible Statistics and their 
Entropy Distance}, Doctoral Dissertation, Friedrich-Alexander University
Erlangen-N\"urnberg (2010)
%
\bibitem{Weis-support} Weis, S.: 
{\it Quantum Convex Support},
Lin.\ Alg.\ Appl.\ 435, 3168--3188 (2011);
correction:\ 436, xvi (2012)
%
\bibitem{Weis-cone} Weis, S.: 
{\it A Note on Touching Cones and Faces},
Journal of Convex Analysis 19, 323--353 (2012)
%
\bibitem{Weis-Knauf} Weis, S.\ and Knauf, A.:
{\it Entropy Distance:\ New Quantum Phenomena},
J.\ Math.\ Phys.\ 53, 102206 (2012) 
%
\bibitem{Weis-topology} Weis, S.:
{\it Information Topologies on Non-Commutative State Spaces},
Journal of Convex Analysis 21 (in press)
%
\bibitem{Wichmann} Wichmann, E.\,H.:
{\it Density Matrices Arising from Incomplete Measurements},
J.\ Math.\ Phys.\ 4, 884--896 (1963)
%
\end{thebibliography}

\end{document}